\documentclass[twocolumn,numberedappendix,twocolappendix,appendixfloats]{openjournal}
\usepackage{natbib}
\usepackage[dvipsnames]{xcolor}
\usepackage{aas_macros}
\usepackage{amssymb}
\usepackage{amsmath}
\usepackage{hyperref}	%
\hypersetup{colorlinks=true,linkcolor=blue,citecolor=blue,filecolor=blue,urlcolor=blue}
\usepackage[caption=false]{subfig}
\usepackage{xspace}
\usepackage{booktabs}
\usepackage{savesym}
\savesymbol{tablenum}
\usepackage{siunitx}
\usepackage{pifont}
\restoresymbol{SIX}{tablenum}
\usepackage[normalem]{ulem}
\usepackage{physics}

\usepackage{cleveref}

\DeclareSIUnit{\Lsun}{\ensuremath{L_\odot}}
\DeclareSIUnit{\Msun}{\ensuremath{M_\odot}}
\DeclareSIUnit{\Zsun}{\ensuremath{Z_\odot}}
\DeclareSIUnit{\hred}{\ensuremath{\textit{h}}}
\DeclareSIUnit\angstrom{\text {Å}}
\renewcommand{\mp}{\ensuremath{\mathrm{m_p}}}
\newcommand{\Rvir}{\ensuremath{R_\mathrm{vir}}}

\crefname{equation}{Equation}{Equations}
\crefname{table}{Table}{Tables}
\crefname{figure}{Fig.}{Figs.}
\crefname{section}{Section}{Sections}
\crefname{subsection}{Subsection}{Subsections}
\crefname{subsubsection}{Subsubsection}{Subsubsections}
\crefname{appendix}{Appendix}{Appendices}

\newcommand{\HI}{H\,\textsc{i}\xspace}
\newcommand{\HII}{H\,\textsc{ii}\xspace}
\newcommand{\Hmol}{H$_2$\xspace}
\newcommand{\HeI}{He\,\textsc{i}\xspace}
\newcommand{\HeII}{He\,\textsc{ii}\xspace}
\newcommand{\HeIII}{He\,\textsc{iii}\xspace}
\newcommand{\CI}{C\,\textsc{i}\xspace}
\newcommand{\CII}{C\,\textsc{ii}\xspace}
\newcommand{\CIII}{C\,\textsc{iii}\xspace}
\newcommand{\CIV}{C\,\textsc{iv}\xspace}
\newcommand{\CV}{C\,\textsc{v}\xspace}
\newcommand{\CVI}{C\,\textsc{vi}\xspace}
\newcommand{\OI}{O\,\textsc{i}\xspace}
\newcommand{\OII}{O\,\textsc{ii}\xspace}
\newcommand{\OIII}{O\,\textsc{iii}\xspace}
\newcommand{\OIV}{O\,\textsc{iv}\xspace}
\newcommand{\OV}{O\,\textsc{v}\xspace}
\newcommand{\OVI}{O\,\textsc{vi}\xspace}
\newcommand{\OVII}{O\,\textsc{vii}\xspace}
\newcommand{\OVIII}{O\,\textsc{viii}\xspace}
\newcommand{\NI}{N\,\textsc{i}\xspace}
\newcommand{\NII}{N\,\textsc{ii}\xspace}

\newcommand{\NV}{N\,\textsc{v}\xspace}
\newcommand{\NVII}{N\,\textsc{vii}\xspace}
\newcommand{\MgI}{Mg\,\textsc{i}\xspace}
\newcommand{\MgII}{Mg\,\textsc{ii}\xspace}
\newcommand{\MgIII}{Mg\,\textsc{iii}\xspace}
\newcommand{\MgX}{Mg\,\textsc{x}\xspace}
\newcommand{\SiI}{Si\,\textsc{i}\xspace}
\newcommand{\SiII}{Si\,\textsc{ii}\xspace}
\newcommand{\SiIII}{Si\,\textsc{iii}\xspace}
\newcommand{\SiXI}{Si\,\textsc{xi}\xspace}
\newcommand{\ionSI}{S\,\textsc{i}\xspace}

\newcommand{\SIII}{S\,\textsc{iii}\xspace}
\newcommand{\SXI}{S\,\textsc{xi}\xspace}
\newcommand{\FeI}{Fe\,\textsc{i}\xspace}

\newcommand{\FeXI}{Fe\,\textsc{xi}\xspace}
\newcommand{\FeXII}{Fe\,\textsc{xii}\xspace}
\newcommand{\FeXXVII}{Fe\,\textsc{xxvii}\xspace}
\newcommand{\NeI}{Ne\,\textsc{i}\xspace}
\newcommand{\NeII}{Ne\,\textsc{ii}\xspace}
\newcommand{\NeIII}{Ne\,\textsc{iii}\xspace}
\newcommand{\NeVIII}{Ne\,\textsc{viii}\xspace}
\newcommand{\NeX}{Ne\,\textsc{x}\xspace}

\newcommand{\ramses}{\textsc{ramses}\xspace}
\newcommand{\vintergatan}{\textsc{vintergatan}\xspace}
\newcommand{\PIEUVthinSSnolocal}{\texttt{PIE-UVB}\xspace}
\newcommand{\PIEUVthinnolocal}{\texttt{PIE-UVB-no-SS}\xspace}

\begin{document}
\title[]{MEGATRON: the impact of non-equilibrium effects and local radiation fields on the circumgalactic medium at cosmic noon}
\author{\vspace{-15mm}
    Corentin Cadiou$^{1,*}$,
    Harley Katz$^{2,3}$,
    Martin P. Rey$^{4}$,\\
    Oscar Agertz$^5$,
    Jeremy Blaizot$^{6}$,
    Alex J. Cameron$^{7}$,
    Nicholas Choustikov$^{7}$,
    Julien Devriendt$^{7}$,
    Uliana Hauk$^{2}$,
    Gareth C. Jones$^{8,9}$,
    Taysun Kimm$^{10}$,
    Isaac Laseter$^{11}$,
    Sergio Martin-Alvarez$^{12}$,
    Kosei Matsumoto$^{13}$,
    Camilla T. Nyhagen$^5$,
    Autumn Pearce$^{2}$,
    Francisco Rodríguez Montero$^{1,2}$,
    Joki Rosdahl$^{6}$,
    Víctor Rufo Pastor$^5$,
    Mahsa Sanati$^{7}$,
    Aayush Saxena$^{7}$,
    Adrianne Slyz$^{7}$,
    Richard Stiskalek$^{7}$,
    Anatole Storck$^{7}$,
    Wonjae Yee$^{2}$
}

\affiliation{$^{1}$Institut d'Astrophysique de Paris, Sorbonne Université, CNRS, UMR 7095, 98 bis bd Arago, 75014 Paris, France}
\affiliation{$^{2}$Department of Astronomy \& Astrophysics, University of Chicago, 5640 S Ellis Avenue, Chicago, IL 60637, USA}
\affiliation{$^{3}$Kavli Institute for Cosmological Physics, University of Chicago, Chicago IL 60637, USA}
\affiliation{$^{4}$Department of Physics, University of Bath, Claverton Down, Bath, BA2 7AY, UK}
\affiliation{$^{5}$Division of Astrophysics, Department of Physics, Lund University, Box 118, SE-221 00 Lund, Sweden}
\affiliation{$^{6}$Centre de Recherche Astrophysique de Lyon UMR5574, Univ Lyon 1, ENS de Lyon, CNRS, F-69230 Saint-Genis-Laval, France}
\affiliation{$^{7}$Sub-department of Astrophysics, University of Oxford, Keble Road, Oxford OX1 3RH, United Kingdom}
\affiliation{$^{8}$Kavli Institute for Cosmology, University of Cambridge, Madingley Road, Cambridge CB3 0HA, UK}
\affiliation{$^{9}$Cavendish Laboratory, University of Cambridge, 19 JJ Thomson Avenue, Cambridge CB3 0HE, UK}
\affiliation{$^{10}$Department of Astronomy, Yonsei University, 50 Yonsei-ro, Seodaemun-gu, Seoul 03722, Republic of Korea}
\affiliation{$^{11}$Department of Astronomy, University of Wisconsin-Madison, Madison, WI 53706, USA}
\affiliation{$^{12}$Kavli Institute for Particle Astrophysics \& Cosmology (KIPAC), Stanford University, Stanford, CA 94305, USA}
\affiliation{$^{13}$Sterrenkundig Observatorium, Universiteit Gent, Krijgslaan 281 S9, 9000 Gent, Belgium}

\thanks{$^*$E-mail: \href{mailto:cadiou@iap.fr}{cadiou@iap.fr}}

\begin{abstract}
    We present three cosmological radiation-hydrodynamic zoom simulations of the progenitor of a Milky Way-mass galaxy from the MEGATRON suite. The simulations combine on-the-fly radiative transfer with a detailed non-equilibrium thermochemical network (81 ions and molecules), resolving the cold and warm gas in the circumgalactic medium (CGM) on spatial scales down to \SI{20}{pc} and on average $\SI{200}{pc}$ at cosmic noon.
    Comparing our full non-equilibrium calculation with local radiation to traditional post-processed photoionisation equilibrium (PIE) models assuming a uniform UV background (UVB), we find that non-equilibrium physics and local radiation fields fundamentally impact the thermochemistry of the CGM.
    Recombination lags and local radiation anisotropy shift ions away from their PIE+UVB values and modify covering fractions (for example, \HI\ damped Ly$\alpha$ absorbers differ by up to $\sim\SI{40}{\percent}$).
    In addition, a resolution study with cooling-length refinement allows us to double the resolution in the cold and warm CGM gas, reaching \SI{120}{pc} on average.
    When refining on cooling length, the mass of the lightest cold clumps decreases tenfold to $\approx \SI{e4}{\Msun}$, their boundary layers develop sharper ion stratification, and the warm gas is better resolved, boosting the abundance of warm gas tracers such as \CIV{} and \OIII.
    Together, these results demonstrate that non-equilibrium thermochemistry coupled to radiative transfer, combined with physically motivated resolution criteria, is essential to predict circumgalactic absorption and emission signatures and to guide the design of targeted observations with existing and upcoming facilities.
\end{abstract}
\keywords{high-redshift galaxies, ISM, galaxy formation}

\section{Introduction}
The circumgalactic medium (CGM) is a complex, multiphase environment that plays a critical role in galaxy evolution, acting as both a reservoir and conduit for baryons cycling in and out of galaxies \citep{tumlinsonCircumgalacticMedium2017}. Observational evidence, notably from absorption-line spectroscopy of background quasars, reveals the multiphase structure of the CGM: neutral hydrogen and low-ionisation ions such as \MgII and \SiII trace cool ($T\sim\SI{e4}{K}$) gas \citep{bergeronAbsorptionLineSystems1986,petitjeanTruncatedPhotoionizedAutogravitating1992,churchillSpatialKinematicDistributions1996,chenBaryonContentDark2008,steidelStructureKinematicsCircumgalactic2010,matejekSurveyMgII2012,tumlinsonCOSHalosSurveyRationale2013,prochaskaQuasarsProbingQuasars2014,werkCOSHalosSurveyPhysical2014,johnsonExtentChemicallyEnriched2017}, while higher ionisation species (\NV, \OVI, \NeVIII) probe the warm-hot phase \citep[$T\sim\SIrange{3e5}{e6}{K}$]{mulchaeyHighIonizationQuasarAbsorption1996,chenOriginChemicallyEnriched2000,stockeGalaxyEnvironmentVI2006,thomSpaceTelescopeImaging2008,savageMultiphaseAbsorberContaining2011,trippHiddenMassLarge2011,tumlinsonLargeOxygenRichHalos2011,meiringQSOAbsorptionSystems2013,pachatDetectionTwoIntervening2017,burchettCOSAbsorptionSurvey2019}.

In the context of observations, it is generally assumed that low ionisation states (e.g.\ \MgII, \SiII, \CII) are in photoionisation equilibrium (PIE) with a UV background (UVB), intermediate ions (e.g.\ \CIV and \CVI) have contributions from both collisional and photoionisation, while high-ionisation states are in collisional ionisation equilibrium (CIE) \citep{gnatTimedependentIonizationRadiatively2007,oppenheimerNonequilibriumIonizationCooling2013,tumlinsonCircumgalacticMedium2017,roca-fabregaCGMPropertiesVELA2019,strawnDistinguishingPhotoionizedCollisionally2022}. This is because collisional ionisation rates scale with temperature, while astrophysical sources only produce copious amounts of photons up to specific energies (i.e.\ stars typically only ionise atoms below $\sim \SI{54}{eV}$ while AGNs can emit photons with energies of \SI{100}{eV} or more \citep{Feltre2016}).

However, there are clear hints from some observations that non-equilibrium effects and ionisation from a local radiation field are important in certain environments. For example, \cite{Kumar2024} studied a sample of \CIV emitters at $z\sim1$ and found at least one absorber where excess radiation beyond the UVB was required to explain the observed ionisation states. \cite{Werk2016} also found evidence that a local radiation field, beyond what is provided by a UVB, may be required to explain low-redshift \OVI absorbers. Likewise, \cite{Kumar2024} found examples where the temperature measured from line widths disagreed with that predicted by PIE models. Finally, the models of \cite{sameerCloudbycloudMultiphaseInvestigation2024} found that half of their observed warm-hot clouds in 47 galaxies at $z<0.7$ require time-dependent photoionisation calculations to explain the observed ion distributions. Despite the evidence of non-equilibrium physics in the CGM and the potential importance of the local radiation field, equilibrium photoionisation models have remained the dominant means by which absorption-line observations are interpreted \citep[e.g.][]{bergeronAbsorptionLineSystems1986,Prochaska2004,werkCOSHalosSurveyPhysical2014}.

Interpreting observations of the CGM in both absorption and emission remains a key challenge. While equilibrium models are simplistic and readily available, they may not be fully representative of the conditions seen in nature. Moreover, time-dependent or non-equilibrium models may overcome some of the limitations of equilibrium assumptions; however, they require additional assumptions about initial conditions, and, for example, whether they are evolved under isobaric or isochoric conditions \citep[e.g.][]{gnatTimedependentIonizationRadiatively2007,oppenheimerAGNProximityZone2013}. For this reason, numerical simulations that attempt to model the properties of the CGM from first principles and follow both the non-equilibrium physics and local radiation are key for interpreting observations.

Despite growing observational constraints on the properties of the CGM, numerical modelling of the CGM remains challenging \citep[see][for a review]{faucher-giguereKeyPhysicalProcesses2023}. Observations suggest the cold phase is populated by clouds with sub-kpc coherence scales (\citealt{rauchSmallScaleStructureHigh2001,crightonMetalenrichedSubkiloparsecGas2015,sternUniversalDensityStructure2016,lanMgIIAbsorbers2017,zahedyCharacterizingCircumgalacticGas2019,
chenCosmicUltravioletBaryon2023,lopezTransverseCluesKiloparsecscale2024a,shabanSpatiallyResolvedCircumgalactic2025}, although cf. \citealt{rubinGalaxiesProbingGalaxies2018,afruniDirectlyConstrainingSpatial2023,duttaProbingCoherenceMetal2024,shabanSpatiallyResolvedCircumgalactic2025}).
A key ingredient to properly capture this cold phase is sufficient computational elements to resolve these structures spatially and to prevent over-mixing between different gas phases.
Several numerical studies of the CGM in idealized environments have found that the properties of gas in the CGM change substantially with improved resolution \citep{scannapiecoLaunchingColdClouds2015,schneiderHydrodynamicalCouplingMass2017,mandelkerColdFilamentaryAccretion2018}, with cold gas fragmenting down to a few \si{pc} or smaller \citep{mccourtCharacteristicScaleCold2018}. Similar results have been found in cosmological simulations \citep{peeplesFiguringOutGas2019,vandevoortCosmologicalSimulationsCircumgalactic2019,hummelsImpactEnhancedHalo2019,rameshZoomingCircumgalacticMedium2024}; however, due to the added complexity of modelling a full cosmological environment, state-of-the-art simulations still only reach resolutions of $\gtrsim\SI{100}{pc}$ in the CGM.

Among the efforts that have been made to improve the resolution in the CGM, the dominant approach has been to uniformly increase resolution in a fixed region surrounding haloes. These studies confirmed that the amount of cold gas increases as resolution improves  \citep{peeplesFiguringOutGas2019,hummelsImpactEnhancedHalo2019,vandevoortCosmologicalSimulationsCircumgalactic2019,rameshZoomingCircumgalacticMedium2024} as does the number of low-mass cold gas clouds \citep{rameshZoomingCircumgalacticMedium2025}.
While effective, this approach has the major drawback of being computationally expensive\footnote{Recall that the volume of the CGM enclosed within $0.2 \Rvir <r<\Rvir$ is \num{125} times larger than that of the ISM.} and limits the complexity of the implemented physics.
Alternatively, some studies proposed to boost resolution in thermally-unstable gas instead (\citealt{reyBoostingGalacticOutflows2024}, see also the second generation of \textsc{foggie} simulations, \citealt{Simons2020}) or on shocks \citep{bennettResolvingShocksFilaments2020}.
Similarly, they found an increase in the cold gas mass fraction in the CGM as well as a two-fold increase in the mass loading factors of supernova-driven outflows in the CGM.
Another explored route was to resolve ionisation fronts \citep{rosdahlExtendedLyaEmission2012a}.

Mass and spatial resolution represent only one aspect where improvement is needed in numerical simulations of the CGM. Beyond resolution, the structure of the CGM is shaped by the interplay between cosmological inflows and galactic outflows, cooling, heating, turbulence, and ionisation. Beginning with gas dynamics, the coupling between inflows and outflows remains highly non-trivial due to its dependence on feedback processes. Recent simulations have, for example, highlighted that the metal content in the CGM is a sensitive probe of feedback models (stellar feedback \citealt{strawnAGORAHighresolutionGalaxy2024,reyARCHITECTSImpactSubgrid2025}, and also AGN feedback, \citealt{oppenheimerAGNProximityZone2013,obrejaAGNRadiationImprints2024,zhangTestingAGNUnified2025}). More recently, simulations are beginning to include cosmic rays that tend to drive much cooler outflows and alter mass outflow rates compared to simulations with only energetic feedback from stars and AGN \citep[see e.g.][]{sharmaThermalInstabilityAnisotropic2010a,dashyanCosmicRAyFeedback2020,girichidisSpectrallyResolvedCosmic2024,rodriguezmonteroImpactCosmicRays2024,kjellgrenDynamicalImpactCosmic2025,kimmImpactCosmicRaydriven2025,weberCRexitHowDifferent2025}. Because of the difference in thermodynamic properties of cosmic ray-driven outflows, the ionisation states in the CGM are modified, leading to, for example, an increase in \OVI and \CIV absorption \citep[see e.g.][for recent simulations]{defelippisEffectCosmicRays2024,thomasWhyAreThermally2025}.

Modelling the chemistry and ionisation states of gas in the CGM is paramount for CGM studies as they directly impact the heating and cooling. Most large-scale simulations typically assume that the CGM is either in collisional ionisation equilibrium (CIE) or photoionisation equilibrium (PIE) with an evolving (but spatially uniform) UV background \citep[e.g.][]{Dubois2014,Schaye2015,Pillepich2018,Dave2019}. Under these assumptions, cooling, heating, and ionisation fractions can be readily tabulated with, for example, spectral synthesis codes such as \textsc{cloudy} \citep{ferland2013ReleaseCloudy2013,ferland2017ReleaseCloudy2017,chatzikos2023ReleaseCloudy2023}, and interpolated on a particle or cell basis (see also \citealt{Sutherland1993}). While some simulations assume a single metallicity scalar and heating and cooling rates are estimated by scaling with solar abundance patterns \citep[e.g.][]{peeplesFiguringOutGas2019,hummelsImpactEnhancedHalo2019,mitchellTracingSimulatedHighredshift2021,reyARCHITECTSImpactSubgrid2025}, a potentially more accurate approach is to follow the enrichment of individual chemical species from various stellar evolutionary processes and model heating and cooling on an element-by-element basis \citep[e.g.][]{wiersmaEffectPhotoionizationCooling2009,Pillepich2018,Schaye2015,Hopkins2018,nelsonFirstResultsTNG502019,Dave2019,Schaye2025}. However, large uncertainties remain on chemical yields \citep[e.g.][]{Buck2021}.

While models that track individual elements are relatively common, few models of the CGM exist that track the individual, non-equilibrium ionisation states of primordial species, metals and molecules. Newer publicly available codes such as \textsc{grackle} \citep{smithGrackleChemistryCooling2017}, \textsc{krome} \citep{grassiKROMEPackageEmbed2014a} or \textsc{chimes} \citep{Richings2014}, have the potential to model this non-equilibrium. However, due to their computational expense, most simulations that consider non-equilibrium physics consider only primordial species, H and He \citep{rosdahlSPHINXCosmologicalSimulations2018,mitchellTracingSimulatedHighredshift2021,reyARCHITECTSImpactSubgrid2025}, while some also include H$_2$ \citep{Schaye2025}. Notable exceptions are \cite{oppenheimerNonequilibriumIonizationCooling2013,oppenheimerBimodalityLowredshiftCircumgalactic2016,oppenheimerMultiphaseCircumgalacticMedium2018} where a detailed non-equilibrium chemistry network of primordial species and metals coupled to a UVB or simple approximations for a local radiation field can be followed for gas at $T>10^4$~K.

Such simulations with complex non-equilibrium chemistry are clearly a small minority among numerical studies of the CGM. For this reason, almost all existing simulations must be post-processed with tools that can predict the ionisation states of individual species in order to be compared with observations. Photoionisation codes such as \textsc{cloudy} \citep{ferland2013ReleaseCloudy2013,ferland2017ReleaseCloudy2017} are commonly used for this purpose \citep[e.g.][]{Shen2013,Nelson2018,peeplesFiguringOutGas2019,hummelsImpactEnhancedHalo2019,Wijers2020}. Likewise, tools such as \textsc{trident} \citep{hummelsTridentUniversalTool2017} wrap tables of these photoionisation models, making them readily available.

As with the models for gas cooling in simulations when individual ionisation states are not followed, when post-processing simulations to compare with observations, some form of equilibrium assumption (e.g.\ PIE, CIE) must be adopted. A common choice is to interpolate PIE metal ionisation states over gas density, temperature, metallicity, and UVB strength (via redshift), under the assumption of a spatially uniform, optically thin UVB. The validity of this equilibrium assumption, particularly in the case of highly metal-enriched gas, is questionable as theoretical work shows that warm gas ($\SI{e4}{K} \lesssim T \lesssim \SI{5e6}{K}$) can have cooling timescales shorter than the recombination timescale, leading to over-ionised gas compared to the equilibrium expectation given the ambient temperature \citep{gnatTimedependentIonizationRadiatively2007,oppenheimerNonequilibriumIonizationCooling2013,Kumar2025}. For gas to reach these non-equilibrium states, some physical mechanism is required to drive the gas out of equilibrium. Such mechanisms include rapid fluctuations of the radiation field \citep{oppenheimerBimodalityLowredshiftCircumgalactic2016}, for example, powered by an AGN \citep{oppenheimerAGNProximityZone2013}, or a high specific star formation rate, or hydrodynamical shocks \citep{kleinHydrodynamicInteractionShock1994} driven, for example, by gaseous inflows or galactic feedback processes (see \citealt{tumlinsonCircumgalacticMedium2017} for a summary of other physics that can lead to non-equilibrium effects).

Importantly, most state-of-the-art cosmological simulations of the CGM do not model the local radiation field around galaxies \citep[e.g.][]{peeplesFiguringOutGas2019,hummelsImpactEnhancedHalo2019,vandevoortCosmologicalSimulationsCircumgalactic2019,rameshZoomingCircumgalacticMedium2024}, and the ones that do \citep[e.g.][]{rosdahlExtendedLyaEmission2012a,mitchellTracingSimulatedHighredshift2021,reyARCHITECTSImpactSubgrid2025} consider only photons with $E>13.6$~eV because the radiation is coupled only to H and He. These sub-ionising photons are key for modelling the gas thermochemistry and ionisation states of cold clouds that give rise to low ionisation state absorbers such as \MgII. The importance of the local radiation field on the properties of the CGM likely depends on the environment and the characteristics of the host galaxy, but in general, it cannot be neglected \citep[e.g.][]{rosdahlExtendedLyaEmission2012a,oppenheimerMultiphaseCircumgalacticMedium2018,Obreja2019,Zhu2024}.

With the recent development of codes such as \textsc{ramses-rtz} \citep{katzRAMSESRTZNonequilibriumMetal2022} and \textsc{sparcs} \citep{chanSPARCSCombiningRadiation2025}, it is now possible to run simulations in which the local radiation from stars and AGN is self-consistently coupled to detailed non-equilibrium thermochemistry networks. Such codes allow one to overcome key limitations of previous simulations that rely on more simplistic equilibrium models both for cooling and for computing ionisation states in post-processing.

\begin{figure*}
    \centering
    \includegraphics[width=.9\textwidth]{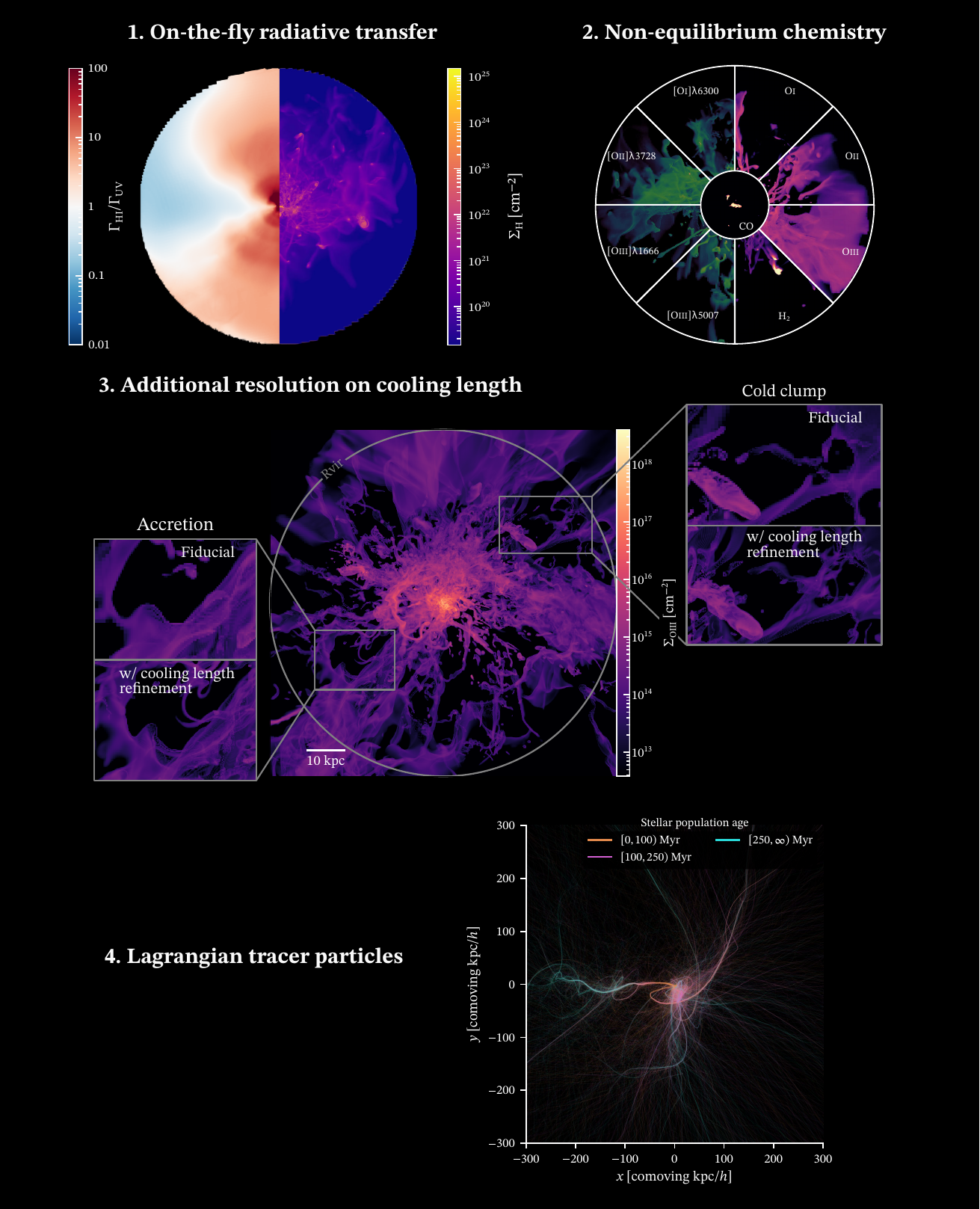}
    \caption{
        We illustrate here the main features of our simulation at $z=4$.
        \textbf{1.} Ratio between the local \HI{}-ionising flux (tracked on the fly) compared to the UV background. The local radiation dominates in a cone above and below the galactic disk.
        \textbf{2.} The radiation field is coupled live to the non-equilibrium thermochemistry of $\geq 80$ species, allowing us to model self-consistently the state of CGM in absorption (e.g., \Hmol, CO and \OI to \OIII column densities), and emission (shown here, [\OI{}]$\lambda\SI{6300}{Å}$, [\OII{}]$\lambda\SI{3728}{Å}$, [\OIII{}]$\lambda\SI{1666}{Å}$ and [\OIII{}]$\lambda\SI{5007}{Å}$ emission maps).
        \textbf{3.} Our fiducial run reaches $\sim\SI{150}{pc}$ in the cold and warm phases ($T<\SI{e5}{K}$) of the CGM.
        Further refining on the cooling length over \SI{150}{Myr}, we reach $\sim\SI{100}{pc}$ in the cold and warm phases and improve the definition of accretion flows and cold clumps
        \textbf{4.} We include tracer particles to follow the Lagrangian evolution of the baryons and understand inflows and outflows in the CGM. We show the past trajectory of gas that formed three populations of stars in the main galaxy.
    }\label{fig:hero}
\end{figure*}

Here, we present early results from the MEGATRON suite of zoom-in cosmological simulations, focusing on the runs designed to study the CGM at cosmic noon. The main features of the simulations are illustrated in \cref{fig:hero}; they aim to address the aforementioned complexities of studying the CGM numerically by combining:\\
\textbf{(i) on-the-fly radiative transfer} sourced by stars on top of an external UV background,\\
\textbf{(ii) a detailed non-equilibrium thermochemical network} including 81 ions and molecules that dominate the thermodynamics,\\
\textbf{(iii)} a fiducial resolution in the cold and warm phase of the CGM down to \SI{20}{pc} (\textbf{$\sim \SI{250}{pc}$} on average) with an increased-resolution simulation refining the CGM (and the IGM), based on cooling length, down to \textbf{$\sim \SI{130}{pc}$} on average, \\
\textbf{(iv) Lagrangian tracer particles} to follow the history of individual gas parcels throughout the baryon cycle.

This setup enables us to study the multiphase structure of the CGM, the cosmic baryon cycle, and synthetic emission and absorption, in a cosmological setting, with unprecedented physical fidelity.
In this paper, we highlight the non-equilibrium physics of the CGM and the importance of the local radiation field that is uniquely captured by our simulations.
Throughout the paper, we adopt the convention that the CGM is defined as the region within $0.2 < r/\Rvir < 1$\footnote{This definition is arbitrary, but is in line with previous work ($0.3-1\,\Rvir$, \citealt{Churchill2013,kocjanHotGasAccretion2024}; $0.2-1\,R_\mathrm{vir}$, \citealt{mitchellTracingSimulatedHighredshift2021}; $>\SI{10}{kpc}$, \citealt{hummelsImpactEnhancedHalo2019}).}.
Our results are qualitatively insensitive to this arbitrary choice.
In particular, we address the question of how far from equilibrium the different phases of the CGM can be, the physics that drives the non-equilibrium behaviour, and how non-equilibrium effects and a local radiation field physically manifest in CGM properties.
Second, we investigate the impact of increased resolution on the ionisation structure of the CGM.

This paper is organised as follows. We present our numerical setup in Section~\ref{sec:numerical_methods}.
We discuss in Section~\ref{sec:non_equilibrium_chemistry} how the structure of the CGM deviates from common assumptions of photoionisation equilibrium.
In Section~\ref{sec:CL_ref}, we explore how the cold gas and ionisation structure respond to increasing the density.
We discuss our results and conclude in Section~\ref{sec:conclusion}.

\begin{figure*}
    \includegraphics[width=\linewidth]{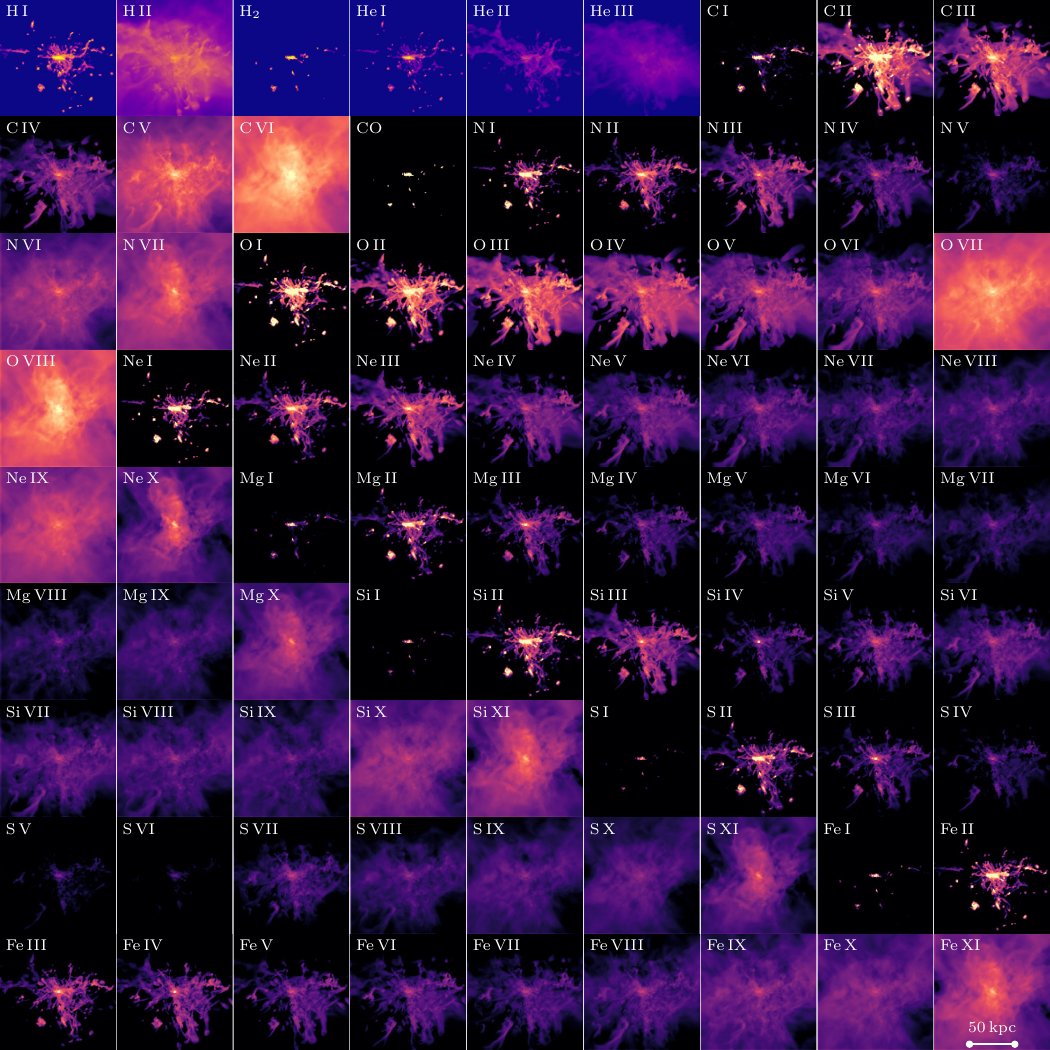}
    \caption{
        Column densities of all the species tracked in the simulations at $z=3$.
        The colormap spans \SIrange{e18}{e22}{cm^{-2}} for primordial species (incl.\ \Hmol), and \SIrange{e12}{e16}{cm^{-2}} for metals and CO.
        The spatial scale is \SI{100}{kpc} on a side (see bottom right).
    }\label{fig:all_ions_column_densities}
\end{figure*}

\begin{figure*}
    \includegraphics[width=\linewidth]{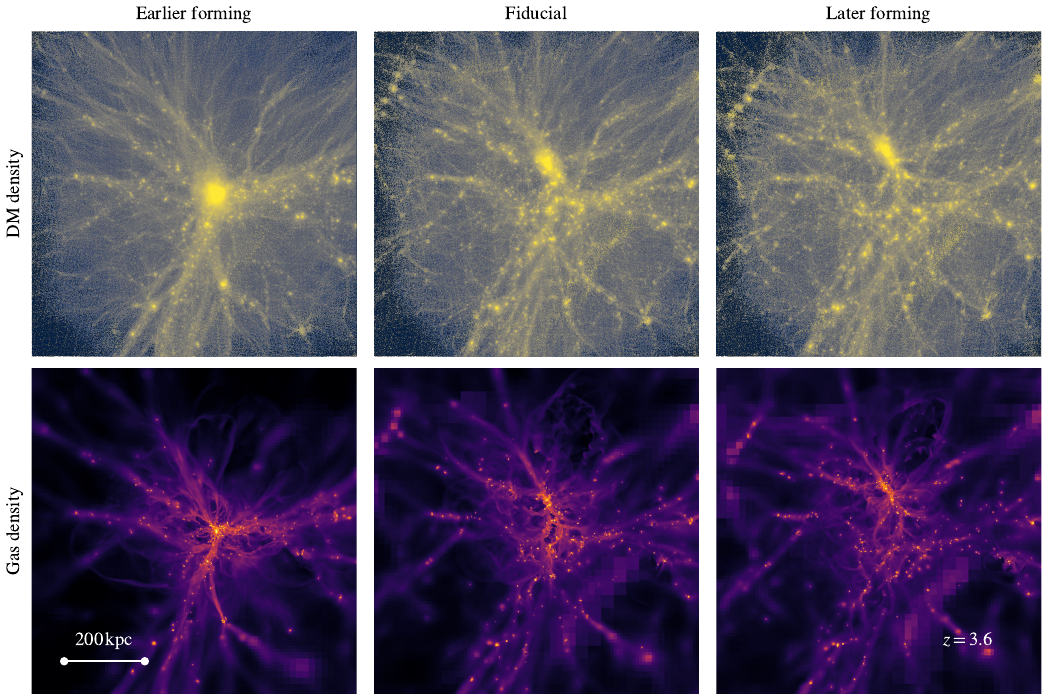}
    \caption{
        Dark matter (left column) and gas density (right column) maps at $z=4$ in our three simulations.
        We `genetically' modify the initial conditions so that the same initial region collapses faster or slower.
        From left to right: the earlier-forming, fiducial, and later-forming scenarios.
        In the earlier-forming scenario, a single halo of mass $\sim \SI{e11}{\Msun}$ is already dominating the region by $z=3.6$, while its formation is delayed in the other two simulations.
        By $z=0$, each of the three regions assembles into a Milky Way-mass halo.
    }\label{fig:three_sims_large_scale}
\end{figure*}

\begin{figure}
  \centering
    \includegraphics[width=\linewidth,trim={0 0 1 1},clip]{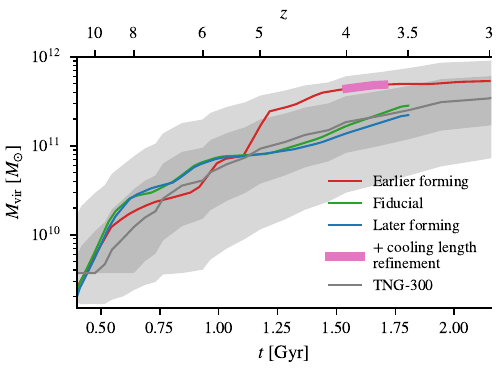}
    \caption{
        Mass assembly histories of the main halo in the three sets of initial conditions compared to the typical mass growth histories of similar mass haloes from IllustrisTNG-300. The dark and light shaded regions represent the $1\sigma$ and $2\sigma$ results from IllustrisTNG-300, with the black line representing the median relation.
        We also include a run with increased resolution for \SI{150}{Myr} using a cooling length criterion, starting at $z=4$.
    }
    \label{fig:mahs}
\end{figure}

\section{Numerical methods}
\label{sec:numerical_methods}

\subsection{Simulation setup}
We present three cosmological radiation hydrodynamical zoom-in simulations of the Lagrangian region of the progenitor of a Milky Way-mass dark matter halo from the MEGATRON suite.
Details of the numerical setup can be found in \cite{Katz2025MegP1}. We summarise here the main features relevant to this work, which are also illustrated in \cref{fig:hero}.

We use the adaptive mesh refinement code \ramses \citep{teyssierCosmologicalHydrodynamicsAdaptive2002} using the \textsc{rtz} thermochemical module with the \textsc{prism ism} model \citep{katzRAMSESRTZNonequilibriumMetal2022,katzPRISMNonEquilibriumMultiphase2022}, as described first in \cite{katzImpactStarFormation2024}. We model multi-frequency radiation hydrodynamics with the following eight energy bins ($[0.1, 1)$, $[1, 5.6)$, $[5.6, 11.2)$, $[11.2, 13.6)$, $[13.6, 15.2)$, $[15.2, 24.59)$, $[24.59, 54.42)$, and $[54.42, \infty)\,\ \si{eV}$) and non-equilibrium chemistry (including all elements contributing at least \SI{.1}{\percent} to the cooling function; $\mathrm{e}^-$, \HI-\HII, \Hmol, \HeI-\HeIII, \CI-\CVI, CO, \NI-\NVII, \OI-\OVII, \NeI-\NeX, \MgI-\MgX, \SiI-\SiXI, \ionSI-\SXI, \FeI-\FeXI).
The chemical network includes ionisation, recombination, and charge exchange processes. Heating and cooling include: photoheating, photoelectric heating, \Hmol formation, \Hmol excitation/dissociation heating, \Hmol cooling, CO cooling, dust recombination cooling, dust-gas collisional processes, primordial cooling, and metal line cooling.
Details of the solved equations and the numerical method can be found in Appendix~\ref{sec:thermochem-model}.
We show in \cref{fig:all_ions_column_densities} column density maps of all species at $z=3$ around the most massive galaxy as viewed edge-on in our suite of simulations.

Star formation occurs in cells above a density threshold ($\rho > \max[\SI{10}{m_p/cm^{3}}, 200\bar\rho]$, where $\bar\rho$ is the background density), where the local turbulent Jeans length is unresolved, at local maxima of the density field, and where the flow is locally convergent \citep{padoanStarFormationRate2011,federrathStarFormationRate2012,kimmFeedbackregulatedStarFormation2017,rosdahlSPHINXCosmologicalSimulations2018}. The number of star particles formed in a timestep follows a Schmidt law \citep{schmidtRateStarFormation1959} where the efficiency per free-fall time depends on gravo-turbulent properties of the gas \citep{padoanStarFormationRate2011,federrathStarFormationRate2012}.
We model feedback from core-collapse supernovæ (SN), type Ia SN, and stellar winds \citep{agertzVINTERGATANOriginsChemically2021}.
Each feedback channel enriches the gas with heavy elements using the yields from \cite{limongiPresupernovaEvolutionExplosive2018,ritterNuGridStellarData2018,seitenzahlThreedimensionalDelayeddetonationModels2013,umedaNucleosynthesisZincIron2002,nomotoNucleosynthesisYieldsCorecollapse2006,nomotoNucleosynthesisStarsChemical2013}.

The Milky Way progenitor is embedded in a cosmological volume of \SI{50}{Mpc/\hred}. Within the zoom-in region, the dark matter (DM) mass resolution is $m_\mathrm{DM,res} = \SI{2.5e4}{\Msun}$ and the minimum mass of each star particle is $m_\mathrm{\star,res}=\SI{3e3}{\Msun}$.
Refinement is only allowed in the Lagrangian region containing all DM particles within $3\Rvir$ of the main halo at $z=0$.
We evolve the simulations with a constant minimum physical resolution\footnote{We allow an additional level of refinement for each doubling of the expansion factor.} of $\approx \SI{20}{pc/\hred}$ down to $z=3$.
We adopt a quasi-Lagrangian refinement strategy, allowing the mesh to be refined whenever $\rho_\mathrm{DM} + \rho_\mathrm{b} \Omega_\mathrm{DM}/ \Omega_\mathrm{b} > 8m_\mathrm{DM,res}/\Delta x^3$, and a Jeans-length refinement criterion, refining cells whenever $4\sqrt{15 k_\mathrm{b} T / (4\pi G \mu \rho_\mathrm{gas})} < \Delta x$, where $\mu$ is the mean molecular weight, $T$ is the temperature and $\rho_\mathrm{gas}$ the gas density.
Compared to our other set of simulations with constant \emph{comoving} resolution \citep{Katz2025MegP1}, we do not model the formation of Pop. III stars, and therefore initialise the metallicity to \SI{e-4}{\Zsun} to account for their metal enrichment \citep{wiseResolvingFormationProtogalaxies2008}.
We employ tracer particles to follow the Lagrangian history of the baryons from accretion to star formation to feedback (\citealt{cadiouAccurateTracerParticles2019}, see also \citealt{genelFollowingFlowTracer2013}).
We use \num{75000000} tracers, each representing a parcel of baryons of mass $m_\mathrm{tracer,res}=\SI{2.5e3}{\Msun}$\footnote{We have on average one tracer per star particle.}. The tracers follow gas fluxes, and can flow from the gas phase into stars (following star formation) and back into the gas (following feedback).

\begin{figure*}
    \includegraphics[width=\linewidth]{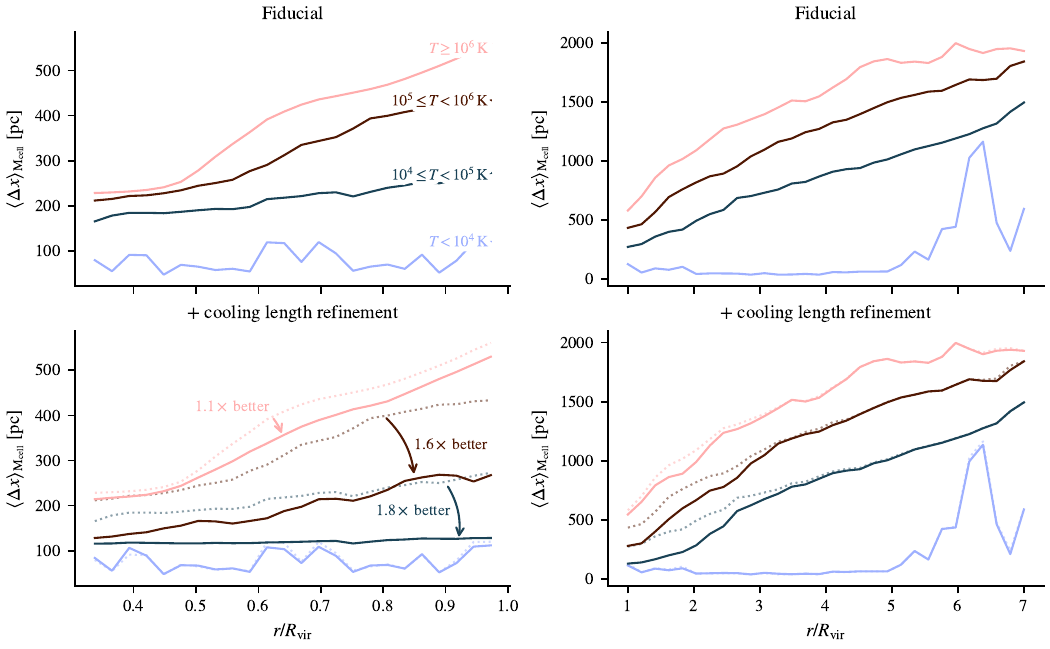}
    \caption{
        Mass-weighted mean spatial resolution for gas at different temperatures in our fiducial simulation (earlier forming) at $z=3.7$ (top row) and when additionally refining on the cooling length (bottom row).
        See the Appendix (\cref{fig:resolution_plot}) for a map of the resolution around the main galaxy at the same redshift.
        Due to the Jeans length refinement, the fiducial run already reaches mean resolutions of the order of \SIrange{100}{150}{pc} in the cold and cool CGM ($T<\SI{e5}{K}$).
        With additional cooling length refinement, the mean resolution is on average better than \SI{130}{pc} or better for all gas except hot gas ($T>\SI{e6}{K}$) in the CGM.
        Focusing on the intergalactic medium (IGM), the cooling length refinement also improves the resolution up to radii of $\sim 3-4 \Rvir$.
    }\label{fig:dx_radial_profile}
\end{figure*}

We perform three simulations with different initial conditions, all starting at $z=99$.
Our initial conditions are based on those of the \vintergatan suite \citep{agertzVINTERGATANOriginsChemically2021,reyVINTERGATANGMCosmologicalImprints2023,joshiPARADIGMProjectHow2025} and were first presented in \cite{reyHowCosmologicalMerger2022}.
We find our stellar mass to agree within \SI{10}{\percent} at $z=4$ with the stellar mass of the \vintergatan simulation, known to be consistent with the $z=0$ stellar mass-halo mass relation \citep{agertzVINTERGATANOriginsChemically2021}.
We employ the genetic modification (GM) technique \citep{rothGeneticallyModifiedHaloes2016} to `velocity-shift' the initial conditions so that the bulk velocity of the main galaxy is as small as possible with respect to the AMR grid, thereby limiting numerical diffusion \citep{pontzenEDGENewApproach2021}.
We also modify the variance of the initial density field \citep{reyQuadraticGeneticModifications2018} to make the main halo form either earlier or later: the effect of these modifications on the large-scale distribution can be appreciated in \cref{fig:three_sims_large_scale}.
To generate the initial conditions, we employ the \textsc{genetic} code \citep{stopyraGenetICNewInitial2021,pontzenPynbodyGenetICVersion2024}.
The effect on the mass assembly history is shown in \cref{fig:mahs}.
This set of three simulations with correlated initial conditions is key to understanding how our control variable -- the assembly time -- affects observable properties of Milky Way-like galaxies at $z<6$.
This provides us with three sets of correlated initial conditions:
these variations in mass assembly history allow us to study their impact on the circumgalactic medium (CGM) structure within the same large cosmological volume.

\subsection{Better resolving the cooling length in the CGM}

One of the main challenges in modelling the CGM is to capture its dynamic range of scales. %
This notably requires sufficient resolution to capture the fragmentation of the gas as it moves from a hot, diffuse phase to a cold, dense phase.

While one approach would be to increase uniformly the resolution in the CGM \citep{vandevoortCosmologicalSimulationsCircumgalactic2019,hummelsImpactEnhancedHalo2019,peeplesFiguringOutGas2019}, we instead focus computational effort on the regions where the gas is expected to cool and fragment. Because of the refinement on the Jeans length, our simulations are already super-Lagrangian and well-capture gravitationally unstable gas. However, to better follow thermally unstable gas, we restart the early-forming simulation at $z=4$ for \SI{160}{Myr} with the cooling length refinement scheme from \cite{reyBoostingGalacticOutflows2024}.
This run is displayed along with the other three simulations in \cref{fig:mahs}.
From the thermal component of the pressure, $P_\mathrm{th}$, the density $\rho$, the temperature $T$, and the net cooling rate, $\Lambda_\mathrm{net}$, we compute the cooling length as
\begin{equation}
    l_\mathrm{cool} = \sqrt{\frac{P_\mathrm{th}}{\rho}} \times \frac{3 \rho k_\mathrm{b} T}{2 \mu m_\mathrm{p}\Lambda_\mathrm{net}}.
\end{equation}
We emphasise here that $\Lambda_\mathrm{net}$ depends on temperature and density, but also on the abundances of each species and importantly on the local intensity of the radiation field (through the heating term and the ionisation state of the gas).
For each cell in the simulation -- including the CGM and IGM -- we trigger additional refinement if the cell size is larger than twice the cooling length.
This is activated for all cells in the high-resolution region and down to a resolution of \SI{120}{pc} (a quarter of the best resolution of the simulation).
As in the fiducial run, cells may be further refined down to \SI{30}{pc} if they are Jeans-unstable or dense enough, as described above.

\begin{figure}
    \includegraphics[width=\columnwidth]{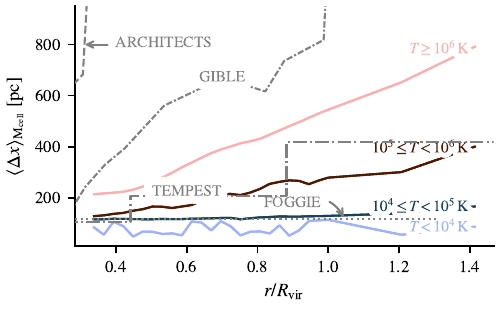}
    \caption{
        Mass-weighted mean spatial resolution for gas at different temperatures when refining on the cooling length (solid lines), shown here at $z=3.7$ and compared to the resolution in the CGM of the \textsc{tempest} \citep[][dotted]{hummelsImpactEnhancedHalo2019},
        \textsc{foggie} \citep[][dash dotted]{peeplesFiguringOutGas2019},
        \cite{vandevoortCosmologicalSimulationsCircumgalactic2019}'s simulations (not visible, since the resolution is \SI{1}{kpc}),
        and the median resolution of the \textsc{gible} simulation \citep[][dash dash dotted]{rameshZoomingCircumgalacticMedium2024} and of the \textsc{architects} simulations \citep[][dashed]{reyARCHITECTSImpactSubgrid2025}.
    }\label{fig:dx_radial_profile_comparison}
\end{figure}
\begin{figure*}
    \includegraphics[width=\linewidth]{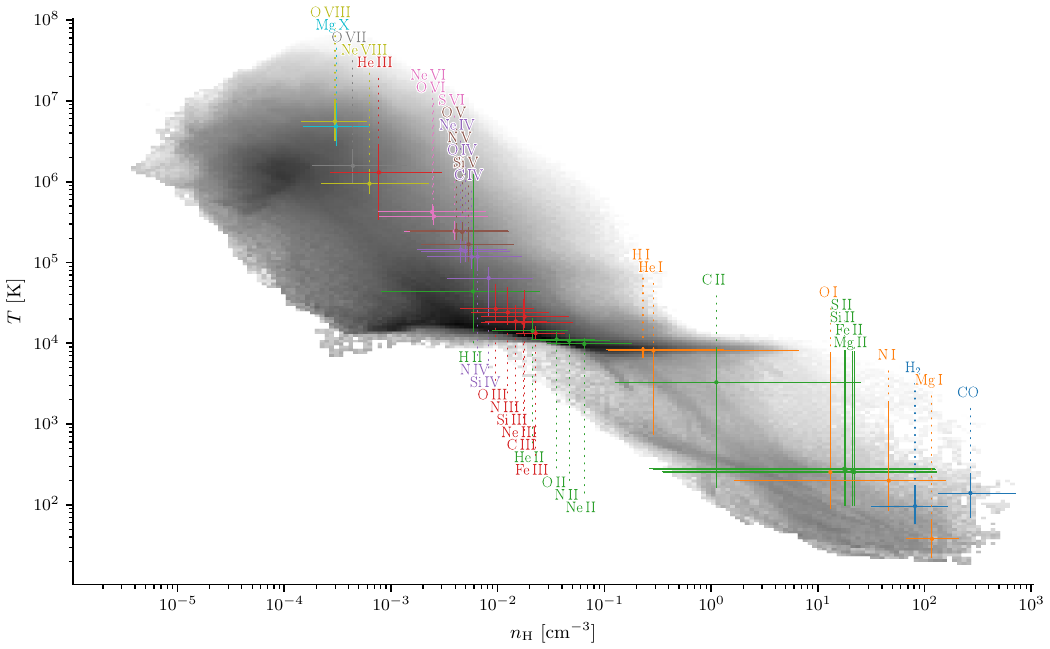}
    \caption{
        Mass-weighted median and [25-75]\si{\percent} percentiles of the temperature and density of different ions in the CGM of the most massive galaxy at $z=3$, excluding gas near any of its subhalo.
        Despite removing gas within $0.2\,\Rvir$ of the main halo or any of its subhalo, we still have a small but significant amount of cold and dense gas, mostly located in the inner CGMs of the main galaxy and its satellites.
        Colours represent the ionisation level.
        See \cite{tumlinsonCircumgalacticMedium2017}, Fig.~6 for a comparison.
        The corresponding ions are annotated above/below the median temperature, from left to right in order of increasing median density.
    }\label{fig:avg_T_vs_avg_nH}
\end{figure*}
\begin{figure*}
    \includegraphics[width=\linewidth,trim={0 {0.05\linewidth} 0 {0.1\linewidth}},clip]{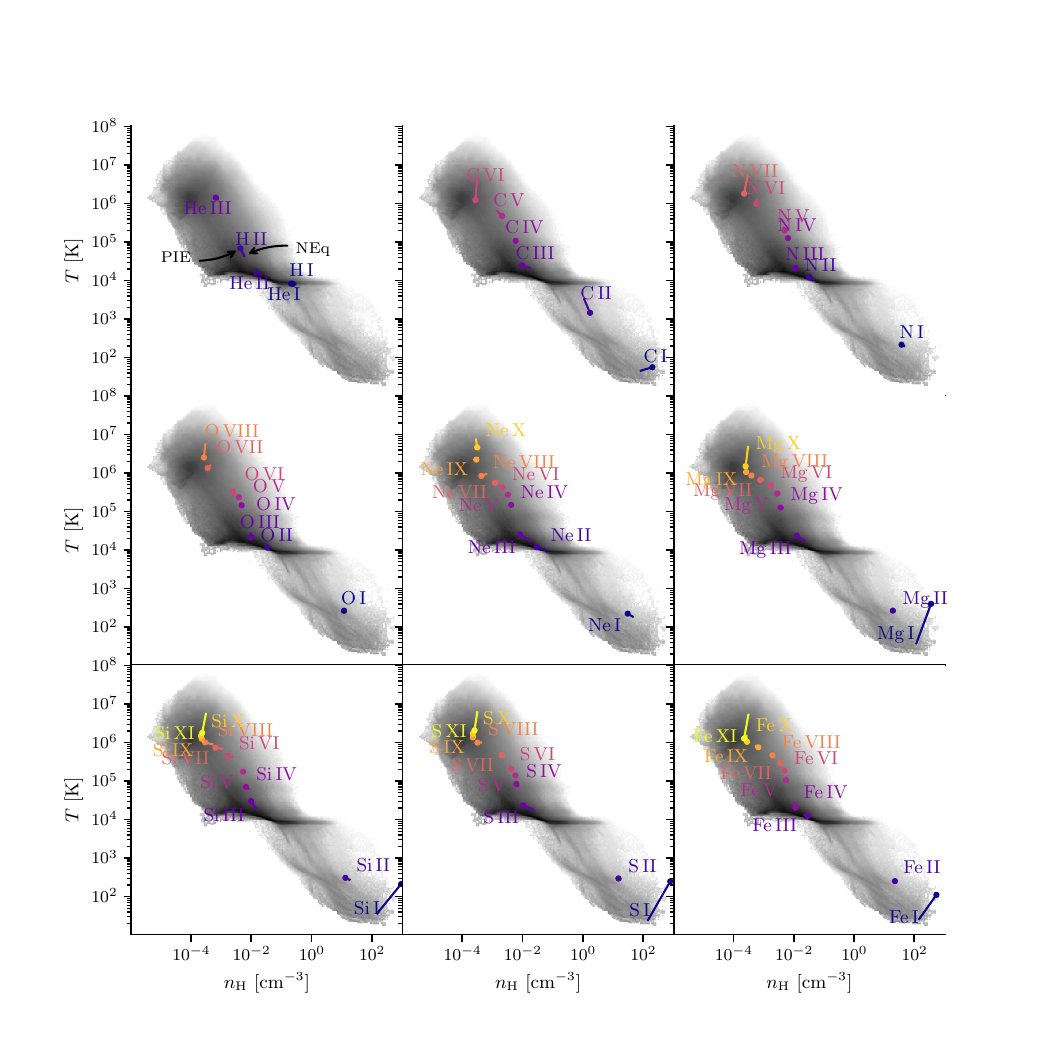}
    \vspace{-1em}
    \caption{
        Difference in the mass-weighted median of the temperature and density traced by different ions in the CGM of the most massive galaxy at $z=4$.
        Relative differences are shown in the Appendix (\cref{fig:avg_T_vs_avg_nH_diff_scatter}).
        The lines start from the value in the non-equilibrium simulation and their tip points to the value assuming \PIEUVthinSSnolocal (dots, abbreviated PIE for brevity).
        Colours represent the ionisation level.
        The corresponding ions are annotated above/below the median density in the non-equilibrium simulation.
        \textbf{As will be made obvious in the following \cref{fig:PIE_vs_NEQ_SS_M_weighted}, large differences in the distribution of gas are not well captured when comparing only the median values.}
        Notwithstanding, we note that some species, such as \CII, \CIII, \NeII and \NeIII, \MgIII, \SiIII, and \SIII display significant differences in their median.
    }\label{fig:avg_T_vs_avg_nH_diff}
\end{figure*}

\begin{figure*}
    \includegraphics[width=\linewidth,trim={0 0 {.1\linewidth} 0},clip]{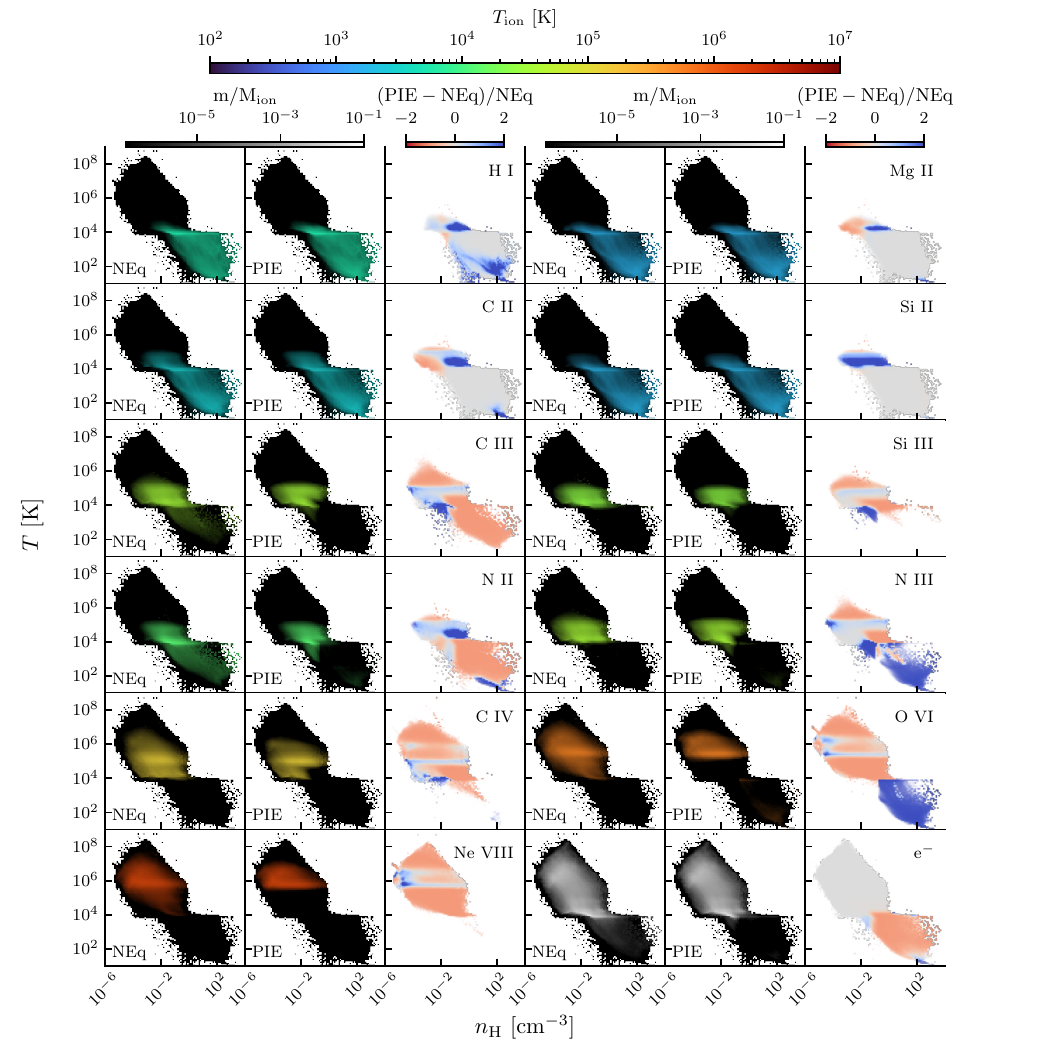}
    \caption{
    Density-temperature diagrams of the CGM for different ions in the non-equilibrium simulation (first and fourth columns), assuming \PIEUVthinSSnolocal in post-processing (second and fifth columns, abbreviated PIE for brevity), and their relative difference (third and sixth columns).
    Each panel is shaded by the mass and coloured proportionally to the mass-weighted temperature (shades of blue for tracers of the cold CGM, shades of red for tracers of the hot CGM).
    \textbf{Significant differences in the amount of gas in the cool-to-warm phases are ubiquitous, with the \PIEUVthinSSnolocal assumption differing from the non-equilibrium simulation by up to a factor of 2 higher or lower.}
}\label{fig:PIE_vs_NEQ_SS_M_weighted}
\end{figure*}

\begin{figure*}
    \includegraphics[width=\linewidth,trim={0 0 0 0},clip]{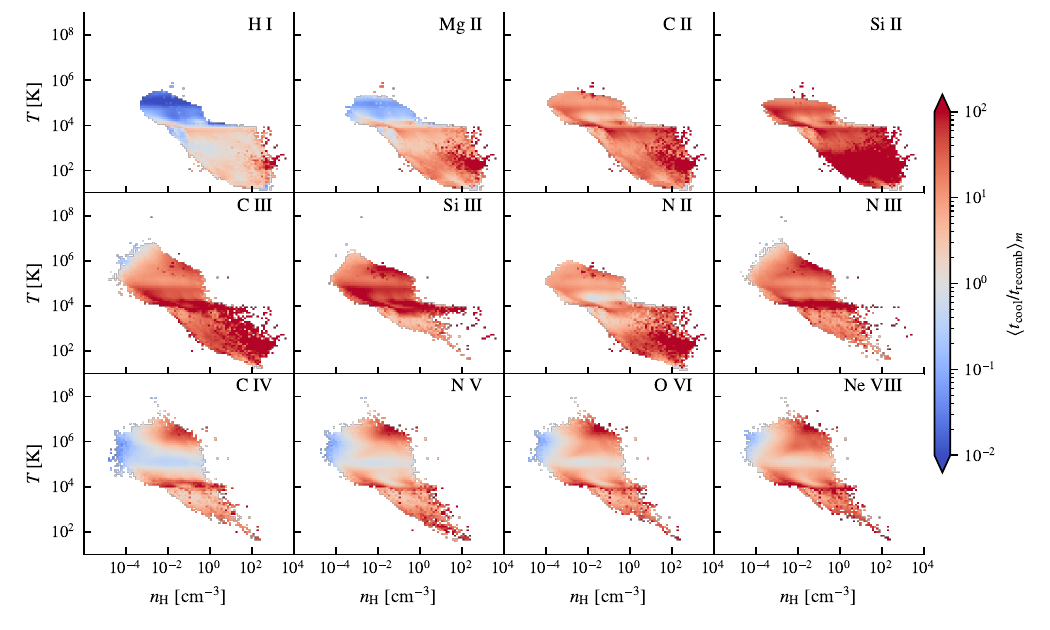}
    \caption{
        Mass-weighted average of the ratio between the recombination timescale and the cooling timescale in the CGM, for the same ions as in \cref{fig:PIE_vs_NEQ_SS_X_weighted}.
        We only include those cells that contribute at least \num{e-6} to the total mass.
        Regions where $t_\mathrm{cool} / t_\mathrm{recomb} < 1$ are more ionised than expected in equilibrium at the local temperature.
        \textbf{Different ions have different recombination timecales, and therefore suffer from recombination lags in different regions of the density-temperature phase space, an effect that can only be captured by non-equilibrium chemistry.}
    }\label{fig:trecomb_tcool_grid}
\end{figure*}

We show in \cref{fig:dx_radial_profile} mass-weighted mean spatial resolution at $z=3.7$ (\SI{100}{Myr} after $z=4$) in different gas phases, and we compare it to other simulations focused on the CGM \citep{vandevoortCosmologicalSimulationsCircumgalactic2019,hummelsImpactEnhancedHalo2019,peeplesFiguringOutGas2019,rameshZoomingCircumgalacticMedium2024,reyARCHITECTSImpactSubgrid2025} in \cref{fig:dx_radial_profile_comparison}.
We also show in the Appendix (\cref{fig:resolution_plot}) a map of the resolution around the main galaxy at the same redshift.
The bump at $r \sim 6-7 \Rvir$ is due to the cool diffuse IGM, and disappears if we remove gas below \SI{e-2}{\mp.cm^{-3}} from the cool phase, where $\mp$ is the proton mass.
In the fiducial run, our resolution in the cold ($T<\SI{e4}{K}$) and cool ($\SI{e4}{K}<T<\SI{e5}{K}$) phases are on average \SIrange{80}{200}{pc}, comparable to the resolution of the \textsc{foggie} simulation.
Once cooling length refinement is activated, resolution in the cold and cool phases reaches an average of \SI{100}{pc}, while the resolution in the warm gas ($\SI{e5}{K} \leq T < \SI{e6}{K}$) increases by a factor of \num{1.6}, reaching $\sim \SI{200}{pc}$ at the virial radius and gradually increasing inwards up to $\sim\SI{100}{pc}$ in the inner CGM.
Overall, the mean resolution in the CGM in the fiducial simulation (resp. in the cooling length simulation) is \SI{200}{pc} (resp. \SI{150}{pc}) at $\Rvir/3$ and \SI{300}{pc} (resp. \SI{200}{pc}) at $\Rvir$.
The quantitative effect of this refinement is discussed in Sec.~\ref{sec:CL_ref}, but can be appreciated qualitatively in \cref{fig:hero}, panel 3. In particular, accretion flows and cold clumps see their resolution significantly improved.

We also emphasise that the resolution in the IGM is also significantly improved, with a mean resolution in the warm gas of \SIrange{200}{1000}{pc}, depending on radius and temperature.
This is to be compared, for example, to TNG-50 \citep{nelsonFirstResultsTNG502019} where the mean resolution at $2\Rvir$ is \SI{5}{kpc}, or to another suite of simulations tailored to study the CGM, the ARCHITECTS zoom-in simulations \citep{reyARCHITECTSImpactSubgrid2025} with typical resolution at $\Rvir$ of \SIrange{1.5}{3}{kpc} (although their simulations reach $z=1$).
As in the CGM, the cooling-length refinement also improves the resolution in the IGM, up to $\sim 3\Rvir$, with a factor of two improvement in the mean resolution at $2\Rvir$.
The effect of this improved resolution on the structure of the IGM will be studied in future work.

An advantage of cooling length refinement over uniform refinement is that it allows us to resolve the CGM at a much lower computational cost: the wallclock time per timestep increases by a factor of two once cooling length refinement is activated.
This allows us to reach similar resolutions in the CGM as the \textsc{foggie} \citep{peeplesFiguringOutGas2019} simulations in gas with $T<\SI{e5}{K}$ (but worse in warmer gas), while giving us exquisite resolutions in the IGM.
The relatively low cost of this additional refinement on cooling length will allow us in the future to extend how long we keep it activated for.
One downside of our approach compared to using a fixed grid is that
changes in resolution artificially convert resolved kinetic energy (in high-resolution cells) into thermal energy (in low-resolution cells), and may thus diffuse away turbulence \citep[see][for a review on the topic]{schmidtLargeEddySimulations2015}.
Both methods have clear advantages; however, the reduced computational cost and number of cells is key for allowing us to fit the large number of ions and molecules into memory.

\subsection{Equilibrium model in the optically-thin limit with a UV background}

In general, the equation describing the evolution of the number density $n_i$ of ion $i$ subject to recombination and collisional ionisation processes, as well as to a photo-ionising UV background is given by
\begin{multline}
    \frac{\mathrm{d} n_i}{\mathrm{d} t} =
        \left[n_\mathrm{e}(\alpha_{i+1} n_{i+1} + \beta_{i-1} n_{i-1}) + \Gamma_{i-1} n_{i-1}f \right] -\\
        \left[n_\mathrm{e}(\alpha_{i} n_{i} + \beta_{i}n_{i}) + \Gamma_{i}n_i f\right],
    \label{eq:ionization_equation}
\end{multline}
where $\alpha_i$ is the recombination rate to a lower state, $\beta_i$ is the collisional ionisation rate to a higher state, and $\Gamma_i$ is the photo-ionisation rate to a higher state.
The parameter $f$ controls whether gas is self-shielded from UV radiation ($f<1$) or not ($f=1$).
In our non-equilibrium simulation, we evolve the equation governing temperature with an equation similar to the one above that additionally includes the effect of local sources of radiation\footnote{This includes notably the absorption of the ionising photons, which means we do not need to assume self-shielding for the local field.}, recombination on dust grains, charge exchange, the formation and destruction of \Hmol and CO, and dust depletion.
This causes the current thermochemical state to depend both on the current temperature and electron number density, but also on the past thermochemical state of the gas and on the local radiation field.

In observations, the past evolution of the thermochemical state and the local radiation fields are typically unknown and \cref{eq:ionization_equation} is usually solved assuming equilibrium ($\mathrm{d}x_i/\mathrm{d}t=0$) yielding the PIE approximation.
Typically, one further assumes an external UV background \citep{haardtRadiativeTransferClumpy2012} and ignores the presence of the unknown local ionising radiation field, and assumes the gas to be optically thin to this background for \HI column densities below \SI{e17}{cm^{-2}} \citep{rahmatiImpactLocalStellar2013}.
In the following, we will refer to this approximation as \PIEUVthinSSnolocal.

As when interpreting observations, the \PIEUVthinSSnolocal approximation is also representative of how one would typically post-process a cosmological simulation to predict metal ionisation states if it were not run with on-the-fly radiation transfer and non-equilibrium metal chemistry\footnote{An extra unknown are the abundances of the different elements, which are typically assumed to be solar when not tracked in the simulation.} as is done in tools such as \textsc{trident} \citep{hummelsTridentUniversalTool2017}.

In order to assess the impact of non-equilibrium chemistry subject to a local radiation field on the structure of the CGM, we post-process our simulations to compute ion abundances assuming \PIEUVthinSSnolocal.
We note that this assumption is one among many (see e.g.\ Section 4.4 of \citealt{katzRAMSESRTZNonequilibriumMetal2022}); our choice is driven by our goal of comparing with the most common assumptions made in the literature.
We generate two datasets: the first assumes self-shielding of \HI gas from the UV background following the same model adopted in the simulation, where we assume the optically thin limit except for species with ionising energy above \SI{13.6}{eV} that are subject to self shielding (\PIEUVthinSSnolocal) and the optically thin limit with no self-shielding (\PIEUVthinnolocal). In both cases, the temperature is fixed to that predicted from the full non-equilibrium thermochemistry model, and we compute the ionisation fractions of all ions (including those not tracked in the non-equilibrium model, such as \FeXII to \FeXXVII).
We emphasise that if we re-ran the simulation assuming \PIEUVthinSSnolocal or \PIEUVthinnolocal, the temperature and density would be different, thereby leading to a different thermodynamical state (with different ion abundances). Similarly, since our cooling function depends on the abundances of individual ions which may themselves have non-solar abundances and may be out of equilibrium, switching to a non-equilibrium model for primordial species only and metallicity-tabulated values for metals (as is done e.g.\ in \textsc{sphinx}, \citealt{rosdahlSPHINXCosmologicalSimulations2018}) would also lead to a different thermodynamical state. Such changes in the state of the CGM would almost certainly alter the star formation history of the galaxy, preventing us from quantitatively disentangling whether the CGM has changed due to non-equilibrium physics or from variations in the star formation history. {\bf For these reasons, the true difference between our full RHD non-equilibrium computation and solutions assuming a UVB and PIE is likely larger than what is presented here.}
For the \PIEUVthinnolocal dataset, we assume $f=1$.
For the \PIEUVthinSSnolocal dataset, we assume ions with an ionisation energy higher than \SI{13.6}{eV} (all but \HI, \CI, \OI, \MgI, \SiI, \ionSI, and \FeI) to be self-shielded from UV radiation using $f = \exp(-n_\mathrm{H} / \SI{e-2}{m_p cm^{-3}})$ \citep[see e.g.][]{faucher-giguereLyaCoolingEmission2010}.
Here, $n_\mathrm{H} = \rho X/m_p$ with $X=0.76$ and $n_\mathrm{He} = \rho (1-X) / 4 \mp$.
We approximate the electron number density as $n_\mathrm{e} = n_\mathrm{H} x_{\rm\HII} + n_\mathrm{He} (x_{\rm\HeII} + 2 x_{\rm\HeIII})$. In fully ionised solar metallicity gas, the maximum the electron fraction can change from the metal contribution is \SI{1}{\percent}.\footnote{In practice, our CGM is at subsolar metallicity and the temperatures are not hot enough to fully ionise the metals, except in low-density cells immediately heated by SN feedback.}
This allows us to compute the equilibrium values regardless of individual metal abundances.

\section{How does the non-equilibrium chemistry impact the ion content of the CGM?}%
\label{sec:non_equilibrium_chemistry}

\begin{figure*}
    \includegraphics[width=\linewidth]{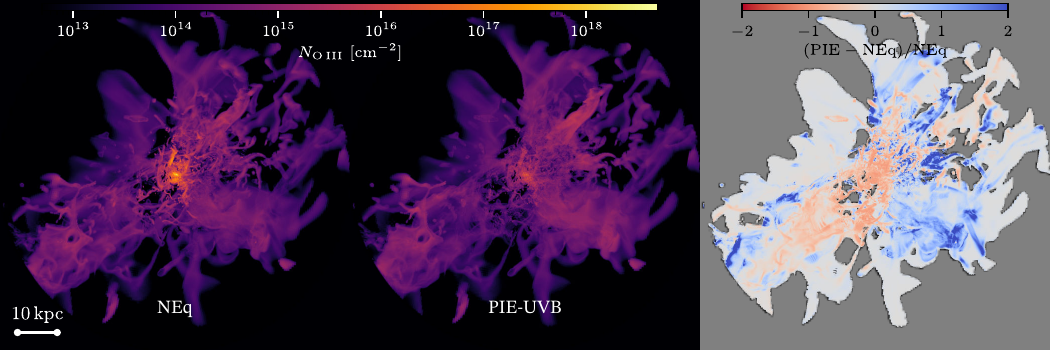}
    \includegraphics[width=\linewidth]{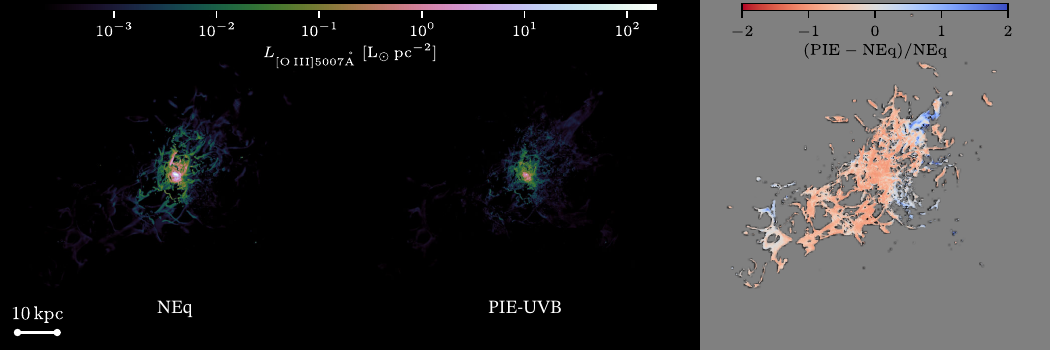}
    \caption{
        \OIII column density (top) and [\OIII{}]\SI{5007}{\AA} surface brightness map (bottom) of the most massive galaxy at $z=4$.
        (Left) Non-equilibrium simulation, (centre) photoionisation equilibrium with self-shielding in the optically thin-limit and neglecting the local ionising field (\PIEUVthinSSnolocal) and (right), the relative difference between the two.
        See \cref{fig:off_axis_O_III_PIE_variations} for the same figure with a different self-shielding prescription and without self-shielding, and \cref{fig:off_axis_H_I_O_I_Mg_II,fig:off_axis_C_IV_O_VI} for a selection of other ions (\HI, \OI, \MgII, \CIV, and \OVI).
        Compared to absorption, the CGM in emission depends additionally on the electron number density and on temperature.
        Consequently, the differences in emission between the non-equilibrium and PIE-SS cases have a different spatial distributions than in absorption.
    }\label{fig:off_axis_O_III_SS}
\end{figure*}

In this Section, we compare the CGM under the typical assumptions of \PIEUVthinSSnolocal to our full non-equilibrium model. First, we show in \cref{fig:avg_T_vs_avg_nH} the mass-weighted median and [25-75]\si{\percent} range of the temperature and density of different ions, \Hmol, and CO in our non-equilibrium model in the CGM of the most massive galaxy in the earlier-forming simulation at $z=4$, illustrating the multiphase nature of the CGM. The ions in the cool, warm, and hot phases ($T>\SI{e4}{K}$) are the same as in Fig.~6 of \cite{tumlinsonCircumgalacticMedium2017}. We have removed any gas located within $0.2\Rvir$ of any halo or subhalo.
We compare in \cref{fig:avg_T_vs_avg_nH_diff} the median temperature and density obtained in our full non-equilibrium model to the values assuming \PIEUVthinSSnolocal (tip of the lines). See Appendix (\cref{fig:avg_T_vs_avg_nH_diff_scatter}) for the relative differences.
While most species appear to be found at similar median temperatures and densities (on a log-log plot), some singly-ionised species have larger differences (especially \CII, \NII, \NeII, and \SiII).
Those species may be ionised by local sources of UV radiation. For those singly-ionised ions, the median temperature and density can shift by a factor of 2.
Neutral species, such as \CI, \MgI, \SiI, \ionSI, and \FeI, which only exist as subdominant species in the coldest and densest parts of the CGM, are almost absent in the \PIEUVthinSSnolocal model, except at the highest densities, which result in large apparent differences. Although these differences for neutral ions are staggering, they only concern species that are not expected to be observable in the CGM.
We find the abundances of \CVI, \NVII, \OVIII, \NeX, \MgX, \SiXI, \SXI, and \FeXII to be lower assuming \PIEUVthinSSnolocal: this is expected since the non-equilibrium model does not include any higher-ionisation states for these elements, while we do assuming in our \PIEUVthinSSnolocal model. Hence this effect is numerical and should be ignored; see discussion in \citealt{katzRAMSESRTZNonequilibriumMetal2022}.

While the median temperature and density traced by most species are within $\pm\SIrange{10}{20}{\percent}$ between the non-equilibrium and \PIEUVthinSSnolocal models, the difference in overall abundance and distribution of those ions in density-temperature phase space can be much greater.
We show in \cref{fig:PIE_vs_NEQ_SS_M_weighted} the density-temperature phase diagrams of different ions in the CGM of the most massive galaxy at $z=4$, shaded by mass fraction (see ionisation-weighted in the Appendix, \cref{fig:PIE_vs_NEQ_SS_X_weighted}) of several ions, comparing the non-equilibrium simulation to the \PIEUVthinSSnolocal assumption.
Here, the overabundance of \CII, \MgII, \SiII can be clearly seen in the densest part of the $T=\SI{e4}{K}$-CGM when assuming \PIEUVthinSSnolocal compared to the non-equilibrium simulation.
Changes to the self-shielding prescription substantially modify the ionisation states in the cold and dense parts of the CGM \citep[as was noted early on][]{fardalCoolingRadiationLya2001,schayePhysicalUpperLimit2001,furlanettoLyaEmissionStructure2005}. However, the differences between the non-equilibrium and PIE models remain qualitatively similar at $T \gtrapprox \SI{e4}{K}$, highlighting that non-equilibrium effects, notably the presence of a local radiation field, are key to set the ionisation structure of the CGM.

\begin{figure}
    \includegraphics[width=\columnwidth]{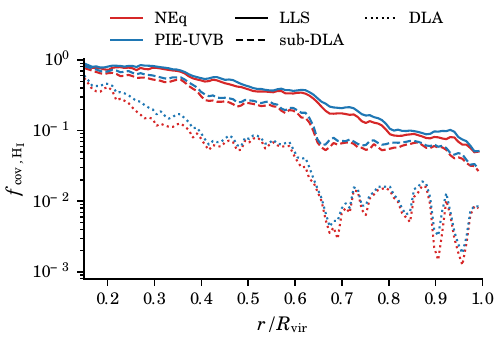}
    \includegraphics[width=\columnwidth]{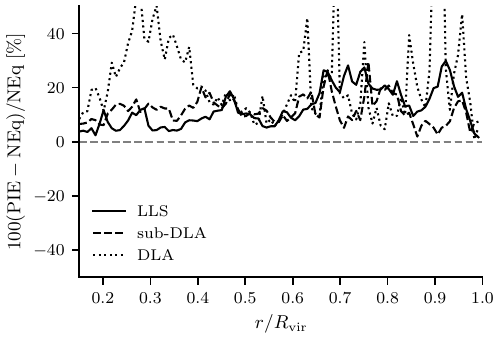}
    \caption{
       (Top) Covering fraction of \HI around the most massive halo at $z=4$ in the non-equilibrium simulation (red) and when assuming the \PIEUVthinSSnolocal.
       (Bottom) Relative difference between the two cases.
       We use here projected (2D) distances.
       Overall, the effect of non-equilibrium chemistry on covering fractions is largest at higher column densities, where \PIEUVthinSSnolocal yields covering fractions larger by up to \SI{40}{\percent} for DLAs.
    }\label{fig:covering_fraction_HI}
\end{figure}

Non-equilibrium effects manifest notably as a recombination lag, where the cooling timescale is shorter than the recombination timescale \citep{kafatosTimeDependentRadiativeCooling1973,gnatTimedependentIonizationRadiatively2007,oppenheimerNonequilibriumIonizationCooling2013,vasilievNonequilibriumCoolingRate2013,katzRAMSESRTZNonequilibriumMetal2022}.
We show in \cref{fig:trecomb_tcool_grid} mass-weighted averages of the ratio between the cooling timescale, defined as
$t_\mathrm{cool} = {3 n k_\mathrm{b} T}/2\Lambda_\mathrm{net}$, and the recombination timescale, defined as $t_{\mathrm{rec},X} = 1 / n_\mathrm{e} \alpha_{X}$, where $\alpha_{X}$ is the recombination rate of ion $X$.
We include here only gas that is cooling efficiently, $t_\mathrm{cool} < \SI{100}{Myr}$, a timescale comparable to the free-fall time of the halo and the time over which star formation varies.

In most regions, we have $t_\mathrm{cool} \gg t_\mathrm{rec}$, i.e.\ chemical equilibrium will be reached faster than thermal equilibrium. Notable low ionisation state exceptions are \HI and \MgII, which show recombination lags at $T \approx \SI{e4}{K}$, $n_\mathrm{H}\sim \SI{e-2}{cm^{-3}}$.
Higher ionisation ions such as \CIV, \NV, and \OVI also display recombination lags close to their equilibrium temperature of \SI{e5}{K}.
The prominence of recombination lag will become more pronounced over time, as metallicity increases, which decreases the cooling timescale while the recombination timescale remains fixed.
While recombination lags allow some ions (\HI, \MgII, and high-ionisation species, such as \CIV, \NV, \OVI and \NeVIII) in the CGM to remain outside of equilibrium, it cannot alone fully account for the non-equilibrium features seen in \cref{fig:PIE_vs_NEQ_SS_M_weighted}. For example, \CIII has clear non-equilibrium features yet its recombination timescale is much shorter than the cooling timescale.
Note that, since the dominant contributor of electrons is hydrogen, any recombination lag on one of its ions will affect indirectly the abundances of other elements through changes to the electron number density.

We have so far focused on the distribution of the different ions in phase space and highlighted the role played by the local radiation field.
As we saw qualitatively in \cref{fig:hero}, panel 1, the local radiation field is highly anisotropic, so we can expect deviations from PIE with a spatially uniform UVB to be anisotropic as well.
Such anisotropic differences may not change the overall abundance of a given ion, nor the temperature and density it traces, but will induce an azimuthal dependence \citep{sameerCloudbycloudMultiphaseInvestigation2024}.
We show in \cref{fig:off_axis_O_III_SS} the \OIII in the CGM as traced in absorption (top row) and in emission (bottom row, [\OIII{}]~$\SI{5007}{Å}$) in the non-equilibrium simulation (left), when assuming \PIEUVthinSSnolocal (centre) and the relative difference between the two (right).
Results when assuming different self-shielding prescriptions can be found in the Appendix (\cref{fig:off_axis_O_III_PIE_variations}), and highlight that self-shielding is critical to correctly map electron number densities and temperature to ionisation fractions.
Additional maps for \HI, \OI, \MgII, \CIV, and \OVI can be found in the Appendix (\cref{fig:off_axis_H_I_O_I_Mg_II,fig:off_axis_C_IV_O_VI}).
Along individual sight lines, the differences can be as high as a factor of two larger or smaller, both in absorption and in emission.
Differences are also anisotropic and not merely column density-dependent, with a spatial distribution that depends on the tracer.
This highlights the importance of considering the local radiation field, which need not be uniform, in addition to the UV background --- a result that is supported both by observations \citep[e.g.][]{Werk2016,Kumar2024} and also in simulations \citep[e.g.][]{oppenheimerAGNProximityZone2013,mitchellTracingSimulatedHighredshift2021,reyARCHITECTSImpactSubgrid2025}.
\begin{figure}
    \includegraphics[width=\linewidth]{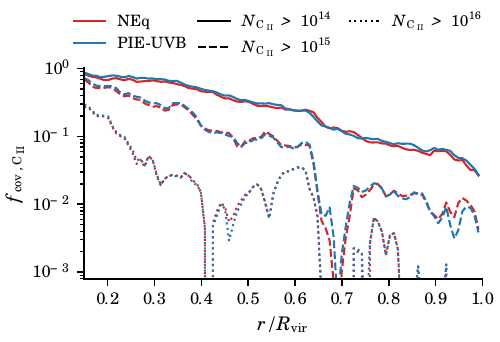} %
    \includegraphics[width=\linewidth]{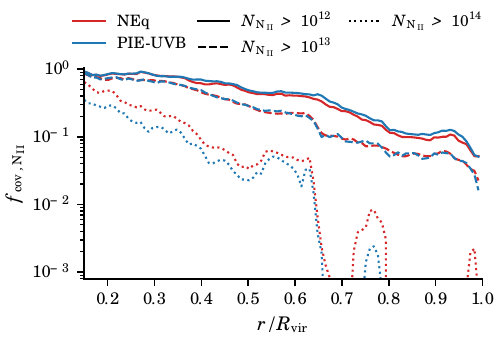} %
    \includegraphics[width=\linewidth]{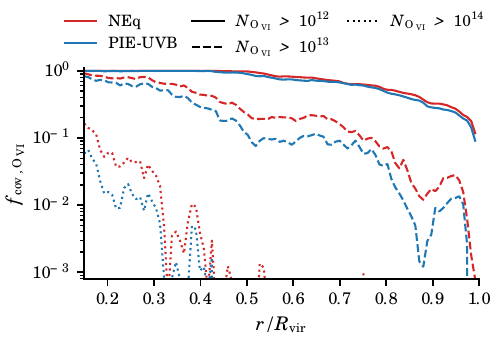} %
    \caption{
        Covering fraction of \CII, \NII, and \OVI, around the most massive halo at $z=4$ in the non-equilibrium simulation (red) and when assuming \PIEUVthinSSnolocal (blue).
        We use here projected (2D) distances.
        Depending on the tracer and column density, covering fractions can change by up to an order of magnitude when assuming \PIEUVthinSSnolocal compared to non-equilibrium.
    }\label{fig:covering_fraction}
\end{figure}

We have seen qualitatively that non-equilibrium effects and local radiation drive anisotropic differences in column density (and emission line surface brightness) maps.
We show in \cref{fig:covering_fraction_HI} how these differences translate into the covering fractions of \HI.
We show the covering fractions in the non-equilibrium simulation (red) and when assuming PIE (blue) for Lyman Limit Systems (LLS, $N_{\rm \HI}>\SI{1.6e17}{cm^{-2}}$), sub-Damped Lyman-$\alpha$ Absorbers (sub-DLAs; $N_{\rm \HI}>\SI{e19}{cm^{-2}}$), and DLAs ($N_{\rm \HI}>\SI{2e20}{cm^{-2}}$).
Assuming PIE leads to a systematic offset of \SIrange{10}{20}{\percent} in the covering fraction of LLS and sub-DLAs, and up to \SI{40}{\percent} for DLAs.
Our results are similar to those obtained in post-processing in \cite{rahmatiImpactLocalStellar2013}.
Note that this effect is not uniform across different ions. For example, looking at a cool gas tracer (e.g., \CII), a warm gas tracer (e.g., \NII) and a hot gas tracer (e.g., \OVI), we show in \cref{fig:covering_fraction} that, for some ions, covering fractions may be smaller when assuming \PIEUVthinSSnolocal at low column densities.

An inspection of the covering fractions of other ions reveals a complex behaviour, with tracers of similar phases either smaller or larger under \PIEUVthinSSnolocal than in the non-equilibrium simulation.

\begin{figure}
    \includegraphics[width=\columnwidth]{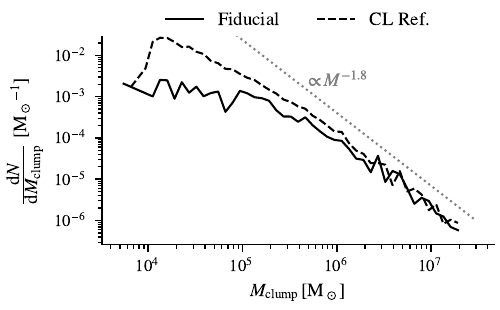}
    \caption{
        Clump mass function without (solid) and with cooling-length refinement (dashed).
        The extra refinement enables us to extend the range of clouds resolved in the simulation by one magnitude.
    }\label{fig:clump_mass_function}
\end{figure}

\begin{figure}
    \includegraphics[width=\columnwidth]{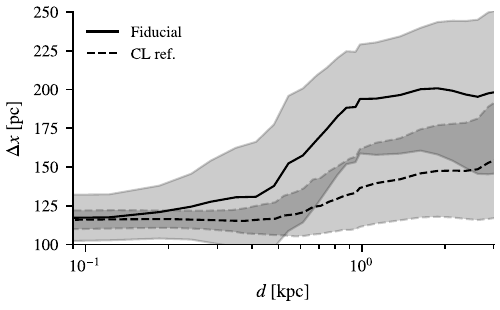}
    \caption{
        Mean 3D radial profile of the resolution in our fiducial (solid) and CGM refinement (dashed) as a function of the distance to the clump ($d=0$ is the clump's edge).
        The cooling length refinement doubles the mean resolution within a few kpc of the clumps, allowing us to better resolve the ionised structure in this region.
    }\label{fig:dx_clump}
\end{figure}

\section{How does increased resolution affect the cold phase of the CGM?}
\label{sec:CL_ref}

Having considered the impact of non-equilibrium effects on the properties of the CGM, in this Section, we study the effect of numerical resolution on the structure of the cold gas in the CGM with and without cooling length refinement.
We focus here on cold clumps, defined as regions of the CGM with a density larger than \SI{e-2}{\mp.cm^{-3}} and a temperature colder than \SI{e4}{K}.
We detect these clumps using a watershed algorithm \citep[based on][]{meyerMorphologicalSegmentation1990} applied to the density field, and retain any clump comprised of at least \num{5} cells and with a maximum density smaller than \SI{1}{\mp.cm^{-3}} to avoid including the ISM of galaxies. We have checked that setting a larger number of cells does not change the results qualitatively, although the mass of the smallest detected clump increases accordingly.

We show in \cref{fig:clump_mass_function} the mass function of the clumps. Similar to \cite{rameshZoomingCircumgalacticMedium2024}, we find that the mass function naturally extends to lower masses as the resolution increases. In particular, we detect cold clumps all the way to \SI{e4}{\Msun} in the cooling length refinement simulation \citep[in agreement with e.g.][]{augustinFOGGIECharacterizingSmallScale2025}, while the fiducial run resolved clumps down to $\sim \SI{e5}{\Msun}$.
We also report no evidence for significant differences in the clump turbulence, as measured by the velocity dispersion within the individual clumps, with increasing resolution.
This suggests that the overall dynamics of the clumps remain unchanged despite the increased resolution.

However, we observe significant differences in the structure of the CGM in and around the clumps.
Indeed, we expect gas close to \SI{e4}{K} and around \SI{e-2}{cm^{-3}} to have short cooling lengths, and hence to be better resolved in the cooling length refinement simulation.
We confirm this in \cref{fig:dx_clump}, where we show the radial profile of the resolution as a function of the distance to the clump -- where $d=0$ is the clump edge -- in the fiducial (solid) and cooling length refinement (dashed) runs.
The cooling length refinement allows us to maintain a uniform resolution of \SI{120}{pc} within \SI{1}{kpc} of the clumps, more than doubling the resolution compared to the (already high) resolution of the fiducial run. This is consistent with the results shown in \cref{fig:dx_radial_profile}, where warm and hot gas ($T<\SI{e6}{K}$) is resolved with $\sim \SI{120}{pc}$.
\begin{figure*}
    \includegraphics[width=\linewidth]{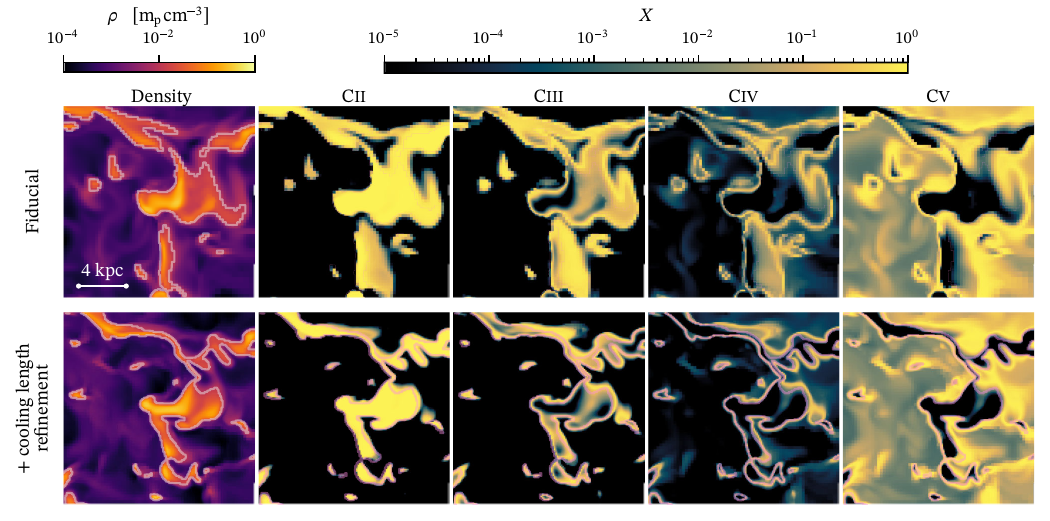}
    \caption{
        Slices of the density field (leftmost column) and the ionisation states of carbon (\CII, \CIII, \CIV, \CV from left to right) around one clump of mass \SI{3e6}{\Msun} at distance \SI{28}{kpc} from the galaxy centre at $z=3.7$.
        The top row shows the clump in the fiducial run, while the bottom row shows the same clump with additional cooling length refinement.
        We highlight the $\rho = \SI{e-2}{\mp.cm^{-3}}$ contour in white and pink.
    }\label{fig:clump_onion_rings}
\end{figure*}
\begin{figure*}
    \includegraphics[width=\linewidth]{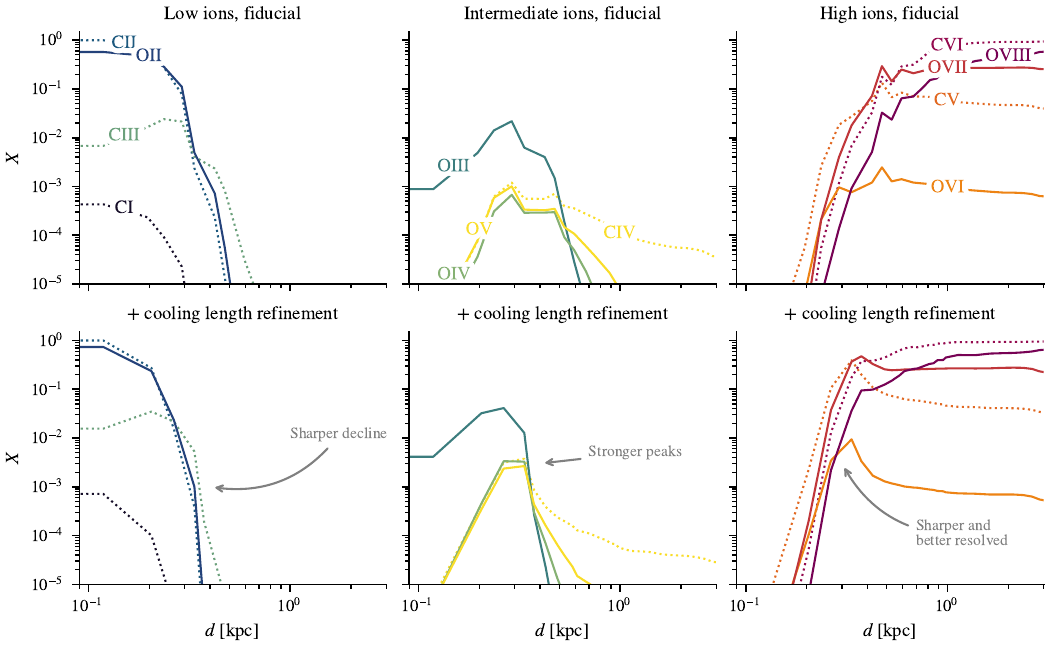}
    \caption{
        Median 3D profile of the carbon (left column) and oxygen (right column) ionisation states around all clumps more massive than \SI{e5}{\Msun} with a maximum density of \SI{1}{\mp.cm^{-3}}.
        We show the profile for the fiducial run (top row) and compare it to the cooling length refinement run (bottom row).
        Additional refinement on the cooling length (dashed line) allows increased resolution in the warm phase, traced by intermediate species living at the boundary of clumps (\OII-\OIII and \CIII-\CIV) to be resolved.
    }\label{fig:clump_ionization_profile}
\end{figure*}

We show in \cref{fig:clump_onion_rings} slices of the ionisation states of carbon around one clump at $z=3.7$ in the fiducial run (top row) and in the cooling length refinement run (bottom row).
The clump is picked as the most anisotropic clump with a mass larger than \SI{e6}{\Msun}.
The density threshold to detect clumps (\SI{e-2}{\mp.cm^{-3}}) is shown as white and pink contours.
We clearly see the `onion-like' structure surrounding the clump, with \CII tracing the cold core, \CIII tracing the boundary of the clump, \CIV tracing its outer interface with the CGM, and \CV tracing the hot CGM.
With additional resolution on the cooling length, \CII and \CIII become less extended with a much sharper outer boundary that almost perfectly tracks the \SI{e-2}{\mp.cm^{-3}} boundary. \CIV traces much more clearly the clump's outer layer (within $\sim \SI{300}{pc}$), and \CV's bump ($\sim \SI{300}{pc}$) away from the clump is also more pronounced, although its structure further away is not changed as dramatically.
Observationally, this translates into a decrease of the $\mathrm{[OIII]}\lambda$\SI{88}{\micro m} luminosity of this specific clump from \SI{6.9e3}{\Lsun} to \SI{5.7e3}{\Lsun} ($-\SI{20}{\percent}$), and $\mathrm{[CII]}\lambda$\SI{158}{\micro m} from \SI{4.4e6}{\Lsun} to \SI{3.8e6}{\Lsun} ($-\SI{15}{\percent}$).
A full analysis of the integrated luminosity across all clumps and all haloes is beyond the scope of the current paper, but, importantly, this highlights that, not only does resolution change the luminosity of individual clouds, but also the relative luminosity of the emissions lines used to probe their structure.
We recover here self-consistently results in qualitative agreement with previous studies that found that the resolution in the CGM \citep{corliesFiguringOutGas2020} and the detailed modelling of the clouds \citep{schimekHighResolutionModelling2024} can significantly affect the luminosity of the associated lines.
Interestingly, our results hint that changes to the resolution affect slightly more $\mathrm{[OIII]}$ than $\mathrm{[CII]}$; if confirmed over the whole CGM, this would be in tension with their results, as they found the latter to be much more sensitive than the former.

\cref{fig:clump_ionization_profile} illustrates these effects quantitatively: it compares the volume-weighted median radial profile of the ionisation states of carbon and oxygen, as a function of the distance to the nearest clump edge, with and without additional refinement\footnote{For readability, we have smoothed the profiles using a Savitzky-Golay filter \citep{savitzkySmoothingDifferentiationData1964} with a window size of 10 and polynomial order 3.}.
Due to the increased resolution in the cooling length refinement simulation, the ionisation structure displays more separation between the ionisation states. For example, the \CII and \OII surrounding the clumps are less extended ($\sim \SI{50}{pc}$ less extended) and drop more sharply in the increased resolution run.

Ions with a median density and temperature close to \SI{e4}{K} and \SI{e-2}{cm^{-3}} (\CIV, \OIII, \OIV and \OV, peaking at \SIrange{200}{400}{pc}) have their maximum abundance enhanced in the increased resolution run -- indicating that they were likely under-resolved in the fiducial run. For example, \OIV peaks at a median ionisation fraction of \num{e-3} in the fiducial run, while it peaks at \num{4e-3} with additional cooling length refinement.
Higher ionisation states (\CV, \OVI and \OVII) that (tentatively) display a bump $\sim \SI{300}{pc}$ away from the clumps edge in the fiducial run have a very clear bump in the cooling length refinement run. Far from the clump, the profiles converge to similar values.

\section{Discussion and Conclusions}%
\label{sec:conclusion}

In this work, we introduced the \textsc{MEGATRON} suite of cosmological radiation-hydrodynamic simulations to investigate the structure and ionisation state of the circumgalactic medium (CGM). Our study was guided by two central questions raised in the introduction: (i) how far from common assumptions of equilibrium are the various phases of the CGM when including both non-equilibrium thermochemistry and a local radiation field? (ii) How does the CGM morphology and ionised structure change with improved spatial resolution on thermally unstable gas?

Our simulation suite is uniquely crafted to study the CGM, combining for the first time the following elements:
\begin{itemize}
    \item \emph{On-the-fly, multi-frequency radiative transfer} from star particles, including both ionising and sub-ionising radiation, as well as a UV background.
    \item \emph{A large non-equilibrium thermochemical network} (81 ions and molecules), designed to follow every ion that contributes significantly to cooling, which allows us to capture recombination lags and non-equilibrium cooling/heating processes self-consistently.
    \item \emph{Cooling-length refinement} and Jeans refinement, ensuring that both gravitationally and thermally unstable regions fragment at the correct physical scale (up to the minimum scale we can resolve in the simulation). This is applied not only inside the virial radius of the main halo but to the entire Lagrange region of the simulation (i.e.\ also the IGM).
    \item \emph{High resolution in the cold and warm phases} ($\lesssim \SI{120}{pc}$), enabling the detection of cold clumps down to $\sim \SI{e4}{\Msun}$.
    \item \emph{Tracer particles} that track Lagrangian baryon histories through inflows, outflows, and recycling, offering a powerful diagnostic of gas flows in and around the CGM.
\end{itemize}

We demonstrated that the inclusion of non-equilibrium (NEq) chemistry coupled with multi-frequency radiative transfer substantially impacts the ionisation structure of the CGM compared to assuming photoionisation equilibrium (PIE) with a spatially uniform UV background and no local radiation field (our \PIEUVthinSSnolocal model).

The combination of non-equilibrium chemistry effects (e.g.\ recombination lags), non-local radiative transfer effects (e.g.\ anisotropic radiation fields) and better-resolved hydrodynamical processes (e.g.\ shocks) allows ions to persist well outside their equilibrium loci in temperature--density space. This translates into systematic offsets in column densities and covering fractions -- for instance, damped Ly$\alpha$ absorbers differ by up to $\sim \SI{40}{\percent}$ between NEq and \PIEUVthinSSnolocal.

We recovered that increasing resolution in the CGM results in an increase of cold gas, and notably an increase in the number of small-mass clouds. Our detailed physical modelling of the thermochemistry allows us to inspect the structure of these clumps, revealing that the increased resolution changes the ionisation profiles around them.
With additional resolution, transitions between different ions are, on average, sharper, and affect their overall luminosity as well as their relative luminosity, notably as observable in $\mathrm{[CII]}$ and $\mathrm{[OIII]}$.

These deviations highlight the necessity of modelling the coupling of non-equilibrium chemistry with local radiation to guide the physical interpretation of CGM absorption and emission data.

\subsection{Discussion}

\textbf{Disentangling between radiative transfer and Neq chemistry:}\\
In this paper, we chose to compare our detailed calculations to more simplistic equilibrium models that are often adopted to interpret observations or to post-process simulations that lack the complex physics included in our work. Such a comparison is known to be imperfect.

For example, several studies have already highlighted that the time-varying and non-local radiation field can significantly affect the ionic structure of the CGM
\citep{sternUniversalDensityStructure2016,sureshOVIAbundanceCircumgalactic2017,oppenheimerFlickeringAGNCan2018,Obreja2019,blaizotSimulatingDiversityShapes2023a,baumschlagerSevenDwarfsIlluminated2025}. Similarly, previous studies computing non-equilibrium chemistry have found, as we report here, that ions residing in the warm-hot phase ($\SI{e4}{K} \lessapprox T \lessapprox \SI{5e6}{K}$, such as \CIV or \OVI) have recombination lags \citep{gnatTimedependentIonizationRadiatively2007,oppenheimerNonequilibriumIonizationCooling2013,Kumar2025}.

The key novelty of \textsc{megatron} is that it allows us to study the explicit coupling between non-equilibrium chemistry for primordial species, metals and molecules, with a self-consistent radiation field in both ionising and sub-ionising bands. For example, unlike \cite{oppenheimerMultiphaseCircumgalacticMedium2018}, we find that these ions are, indeed, pushed away from PIE with a UVB.
Given the numerous differences between simulations, we suggest two possible reasons for our different results: (1) our thermochemical model is fully coupled to the radiation field, which itself can fluctuate on short timescales thanks to the injection of photons by young stars. This is especially true at $z\gtrapprox 3$, where star formation is expected to be much more stochastic than in the low-$z$ Universe.
(2) We have higher resolution in the CGM, which allows the Reynolds number to be much larger, allowing stronger hydrodynamical shocks to develop and preventing mixing of hot gas with colder one that would otherwise dilute high ionisation states.

Looking ahead, \textsc{megatron} is well-suited as a basis for developing efficient forward-modelling schemes for CGM diagnostics that could be applied in post-processing to cosmological simulations.
The tracer particles allow us to follow individual parcels of gas through their full thermodynamical history, capturing how they are steered away from equilibrium.
This is particularly important because, as our results illustrate, knowledge of the instantaneous radiation field would be necessary but not sufficient to construct such a scheme: recombination lags tie the current ionisation state of the gas to its past temperature and ionisation history, neither of which is encoded in any instantaneous quantity.
We leave the construction of such a scheme to future work.

\textbf{Impact of resolution and cold-phase structure:}\\
Our simulation with cooling-length refinement demonstrated that resolving the cooling scale is critical for capturing the multiphase structure of the CGM (see also e.g. \citealt{peeplesFiguringOutGas2019,hummelsImpactEnhancedHalo2019}). Higher resolution (from $\sim \SI{200}{pc}$ down to $\sim \SI{120}{pc}$ in cold/warm gas) produces an order-of-magnitude increase in the number of low-mass cold clumps ($M \sim \SI{e4}{\Msun}$) (see \citealt{rameshZoomingCircumgalacticMedium2024} for qualitatively similar results), sharper ion gradients at clump boundaries, and enhanced abundance of intermediate tracers (e.g. \CIII, \OIII).
These changes to the ionisation gradients convert into changes to the luminosity of the associated tracers, for example $\mathrm{[CII]}$ and $\mathrm{[OIII]}$.
Our results are in agreement with previous studies that showed that line luminosities are sensitive to the detailed modelling of the CGM structure \citep{corliesFiguringOutGas2020,schimekHighResolutionModelling2024}.
The key contribution of our work is to obtain those results with a setup that self-consistently evolves the thermodynamical state of the CGM together with the radiation field, that relaxes the assumption of being in equilibrium, with state-of-the-art resolution.
This supports the picture that small-scale fragmentation and mixing layers -- which we still do not fully resolve yet -- are essential for shaping observable CGM signatures.

\noindent\textbf{Comparison to previous simulations:}\\
In terms of resolution, our simulations are comparable to the \textsc{foggie} simulations \citep{peeplesFiguringOutGas2019} and the \textsc{tempest} simulations \citep{hummelsImpactEnhancedHalo2019}.
Our simulations most closely resemble the SPHINX model \citep{mitchellTracingSimulatedHighredshift2021,reyARCHITECTSImpactSubgrid2025}.
One of the novelties of this work is the self-consistent evolution of a large thermochemical network on-the-fly coupled with yields from supernovae and AGB winds, while SPHINX assumed solar abundances and tabulated cooling rates for metal cooling.
Focusing on the two main contributors to metal cooling, carbon and oxygen, we find that $\mathrm{[C/O]}$ has spatial fluctuations on the order of $\pm 0.3$ in the CGM.
This will for example lead to local fluctuations of a factor of $\sim 2$ at \SI{e5}{K}, where oxygen dominates the cooling function \citep{wiersmaEffectPhotoionizationCooling2009}. Likewise, we also consider radiation with energies $<\SI{13.6}{eV}$. Neither \textsc{foggie} nor \textsc{tempest} include on-the-fly radiation transport, while \cite{mitchellTracingSimulatedHighredshift2021,reyARCHITECTSImpactSubgrid2025} only include ionising radiation.

\textbf{Numerical considerations on the cooling-length refinement approach:}\\
Here, we discuss the relevance of increasing cooling length in light of the two mechanisms proposed by \cite{hummelsImpactEnhancedHalo2019} to explain how increased resolution changes the structure of the CGM.
First, we have shown in \cref{fig:clump_mass_function} that the increased resolution leads to the creation of small-mass cold clumps, while the statistics of larger clumps are left unchanged.
This is consistent with the idea that higher resolution allows for better seeding of precipitation
\citep{2012MNRAS.419.3319M,2012MNRAS.420.3174S,2015Natur.519..203V}.
We argue that running with cooling length refinement for \SI{100}{Myr} is sufficient to capture precipitation.
Indeed, gas at density \SI{2e-4}{cm^{-3}}, temperature \SI{2e5}{K}, and metallicity of $Z=0.05\,Z_\odot$ has a typical cooling timescale on the order of \SI{60}{Myr} \citep{hummelsImpactEnhancedHalo2019}, with the timescale decreasing with increasing (initial) density and decreasing temperature.
These timescales are shorter than the time over which we run with additional refinement (\SI{150}{Myr}), allowing us to properly resolve the formation of clumps at the threshold density we target (\SI{e-2}{cm^{-3}}).
Additional cooling instabilities in the IGM would result in an increase of the total number of cells in the simulation; in our simulation, we do not find evidence for such an increase after a few tens of \si{Myr}, suggesting that all the gas that was cooling-unstable had already been refined.

Second, the cooling length refinement also reduces the artificial numerical mixing between hot and cold phases and allows for a warm phase to develop, as we showed in \cref{fig:clump_ionization_profile}, in line with previous studies \citep{hummelsImpactEnhancedHalo2019,vandevoortCosmologicalSimulationsCircumgalactic2019,peeplesFiguringOutGas2019}.
However, compared to other studies, our additional cooling-length refinement is activated everywhere and not merely within some distance from the galaxy.
This allows us to avoid the artificial mixing caused by resolution jumps when gas is advected from a low- to a high-resolution region, though it comes at the price of having resolution jumps moving from cooling-stable to cooling-unstable regions.
While our additional refinement strategy is designed to improve resolution around advected gas, it may be less optimal to improve turbulence, as the turbulent cascade may be artificially broken at the cooling-length refinement boundary.

At $z=4$, accretion on all galaxies in our simulations is dominated by cold flows \citep{dekelGalaxyBimodalityDue2006}.
Since we allow for refinement on the cooling length to happen in the CGM and in the IGM, our simulation suite is particularly well-suited to study the structure of well-resolved cold flows as they interact with the CGM,
as well as model their observability (in Ly$\alpha$ notably, \citealt{rosdahlExtendedLyaEmission2012a,mitchellTracingSimulatedHighredshift2021}).
Cold flows may entrain hot gas by precipitation on their edges \citep{mandelkerInstabilitySupersonicCold2020,aungEntrainmentHotGas2024a}, provided resolution is sufficient to resolve the mixing layer and the onset of hydrodynamical instabilities, notably the Kelvin-Helmholtz instability.
The presence of eddy-like structure at the edge of filaments is qualitatively apparent in \cref{fig:hero}, panel 3, but should be further investigated.
In particular, we could replicate the approach developed in \cite{mitchellTracingSimulatedHighredshift2021} (we employ the same tracer particles approach) to follow inflows (and outflows) and relate their observable properties to their thermodynamical history.

Our suite of simulations provides a framework to interpret absorption and emission observations of the CGM and IGM around high-redshift galaxies with greater fidelity, and to guide the design of targeted campaigns with upcoming observational facilities.
We, however, acknowledge that key ingredients are still missing from our suite of simulations: we do not model AGN feedback \citep{oppenheimerAGNProximityZone2013,obrejaAGNRadiationImprints2024}, cosmic rays and magnetohydrodynamics \citep{jiPropertiesCircumgalacticMedium2020,defelippisEffectCosmicRays2024,thomasWhyAreThermally2025,kjellgrenDynamicalImpactCosmic2025}, or thermal conduction \citep{sharmaThermalInstabilityAnisotropic2010a,armillottaSurvivalGasClouds2017,talbotAnisotropicThermalConduction2025}, while our results emphasise a lack of convergence of the observable properties with resolution.
Nevertheless, taken together, our results demonstrate that non-equilibrium thermochemistry and physically motivated resolution criteria are necessary for modelling the multiphase CGM, and for correctly interpreting observational data.

\section*{Acknowledgements}
CC would like to thank Hsiao-Wen Chen, Pierre Guillard, and Santi Roca Fàbrega for their valuable feedback.
AS and AJC acknowledge funding from the “FirstGalaxies” Advanced Grant from the European Research Council (ERC) under the European Union’s Horizon 2020 research and innovation programme (Grant agreement No.789056).
AS acknowledges support from the Science and Technology Facilities Council (STFC) for a PhD studentship.
FRM is supported by the Kavli Institute for Cosmological physics at the University of Chicago through an endowment from the Kavli Foundation and its founder Fred Kavli.
GCJ acknowledges support by the Science and Technology Facilities Council (STFC), by the ERC through Advanced Grant 695671 ``QUENCH'', and by the UKRI Frontier Research grant RISEandFALL.
The material in this manuscript is based upon work supported by NASA under award No. 80NSSC25K7009. HK acknowledges support from FACCTS.
KM acknowledges the Flemish Fund for Scientific Research (FWO-Vlaanderen), Grant number 1169822N.
MS acknowledges the support from the Swiss National Science Foundation under Grant No. P500PT\_214488.
NC acknowledges support from the Science and Technology Facilities Council (STFC) for a PhD studentship.
OA acknowledges support from the Knut and Alice Wallenberg Foundation, the Swedish Research Council (grant 2019-04659), the Swedish National Space Agency (SNSA Dnr 2023-00164), and the LMK foundation.
TK is supported by the National Research Foundation of Korea (RS-2022-NR070872 and RS-2025-00516961) and by the Yonsei Fellowship, funded by Lee Youn Jae.

This work has made use of the Infinity Cluster hosted by Institut d'Astrophysique de Paris.
This work made extensive use of the dp265, dp016, dp373, and dp379 projects on the DiRAC ecosystem. This work was performed using the DiRAC Data Intensive service at Leicester, operated by the University of Leicester IT Services, which forms part of the STFC DiRAC HPC Facility (www.dirac.ac.uk). The equipment was funded by BEIS capital funding via STFC capital grants ST/K000373/1 and ST/R002363/1 and STFC DiRAC Operations grant ST/R001014/1. 
This work used the DiRAC@Durham facility managed by the Institute for Computational Cosmology on behalf of the STFC DiRAC HPC Facility (www.dirac.ac.uk). The equipment was funded by BEIS capital funding via STFC capital grants ST/P002293/1, ST/R002371/1 and ST/S002502/1, Durham University and STFC operations grant ST/R000832/1. 
This work was performed using resources provided by the Cambridge Service for Data Driven Discovery (CSD3) operated by the University of Cambridge Research Computing Service (www.csd3.cam.ac.uk), provided by Dell EMC and Intel using Tier-2 funding from the Engineering and Physical Sciences Research Council (capital grant EP/T022159/1), and DiRAC funding from the Science and Technology Facilities Council (www.dirac.ac.uk). DiRAC is part of the National e-Infrastructure.
This work has made use of the Infinity Cluster hosted by Institut d'Astrophysique de Paris. We thank Stephane Rouberol for running smoothly this cluster for us.
The authors thank Jonathan Patterson for smoothly running the Glamdring Cluster hosted by the University of Oxford, where part of the data processing was performed.
The authors also acknowledge financial support from Oriel College’s Research Fund.

This project made use of the following packages:
\textsc{cmasher} \citep{vanderveldenCMasherScientificColormaps2020},
\textsc{gnu parallel} \citep{tange_ole_2018_1146014},
\textsc{joblib} \citep{developersJoblib2025},
\textsc{matplotlib label lines} \citep{cadiouMatplotlibLabelLines2022},
\textsc{numpy} \citep{harrisArrayProgrammingNumPy2020},
\textsc{scipy} \citep{2020SciPy-NMeth},
and
\textsc{yt} \citep{turkYtMulticodeAnalysis2011}.

\bibliographystyle{mnras}
\bibliography{authors_short} %

\appendix
\section{Thermochemistry model}\label{sec:thermochem-model}

We provide in \cref{tab:thermo-defs} the governing equations for the evolution of the number density of ionised species, $\mathrm{CO}$, and $\mathrm{H_2}$, as well as energy, photon density, and photon fluxes.
For brevity, we have omitted in \cref{eq:thermochemistry-n} sink and source terms corresponding to $\mathrm{CO}$ and $\mathrm{H_2}$ formation and destruction (Equations~\ref{eq:thermochemistry-CO} and \ref{eq:thermochemistry-H2}).
Additional information about the equations can be found in the papers that introduced them in \ramses{}, cited alongside the equations of \cref{tab:thermo-defs}.
In our model, all recombination rates and collisional ionisation rates are computed on-the-fly for each ion as functions of temperature. For each photon group, we compute luminosity-weighted ion-by-ion cross sections, where the luminosities are measured from the SEDs of the star particles in the simulation and recomputed every ten coarse timesteps.
These cross sections are then used to compute the local photoionisation rate of ion $i$ that we add to an external photoionisation rate assuming a~\cite{haardtRadiativeTransferClumpy2012} UV background modulated by self-shielding.
We further assume the on-the-spot approximation and thus do not track ionising photons emitted when recombining to the ground state.

This thermochemistry network yields a large (81 species with 8 photon groups in this project) set of coupled ordinary differential equations (ODEs).
We use operator splitting: the hydrodynamics and radiative transfer are evolved at their own (larger) timesteps, $\Delta t_\mathrm{hydro}$ and $\Delta t_\mathrm{RT} \leq \Delta t_\mathrm{hydro}$ respectively, while the chemistry ODE system in each cell is advanced over $\Delta t_\mathrm{RT}$ as a standard initial value problem (IVP), subcycled with an internal adaptive timestep $\Delta t_\mathrm{IVP} \leq \Delta t_\mathrm{RT}$.
For temperature, we employ a backward Euler method, simplifying \cref{eq:thermochemistry-T} by replacing the right hand side with its first-order Taylor expansion--evaluated numerically--in temperature.
For photon density and fluxes, we employ a standard forward Euler method, as the problem is not stiff for those quantities and evolves over timescales that are typically much longer than the cooling and recombination timescales.
Ionisation levels require most care, since they can change over very short timescales and are strongly coupled; we employ a semi-implicit solver, using the updated values of species $i$ to update species $i+1$.
Following~\cite{rosdahlRAMSESRTRadiationHydrodynamics2013}, $\Delta t_\mathrm{IVP}$ is reduced adaptively so that no quantity changes by more than \SI{10}{\percent} per subcycle.
This is the ODE solver presented in \cite{katzRAMSESRTZNonequilibriumMetal2022}, which was shown to be stable, and to reproduce results consistent with Cloudy when evolved to equilibrium.

In photon bins where multiple species have ionisation thresholds (e.g.\ \HI{} and \HeI{} in the $[24.59, 54.42)$ \si{eV} bin), photon conservation is ensured automatically by \cref{eq:thermochemistry-N,eq:thermochemistry-F}: the total absorption rate, $c N_j \sum_i^\mathrm{species} n_i \bar{\sigma}_{i,j}$ exactly equals the sum of individual photoionisation rates. Photons are therefore implicitly split between competing absorbers in proportion to their relative opacity contributions, with no additional approximation required \citep[cf.][Appendix D]{friedrichRadiativeTransferEnergetic2012}.
The accuracy of this approach is ultimately limited by the photon bin widths: luminosity-weighted cross sections reduce the sensitivity to binning, but cannot fully capture spectral features within a bin. Our bins were chosen to straddle the ionisation thresholds of \HI{}, \Hmol{}, \HeI{}, and \HeII{} (at 13.6, 15.2, 24.59, and \SI{54.42}{eV}), ensuring that photons capable of ionising each primordial species are tracked in separate groups. Finer binning would improve accuracy, albeit not very significantly \citep{katzRAMSESRTZNonequilibriumMetal2022}, for ions of other species, but at prohibitive computational cost.

\begin{table*}
  \center
  \caption{%
    Summary of the implemented thermochemical model.
    We provide in \cref{eq:thermochemistry-e,eq:thermochemistry-F,eq:thermochemistry-CO,eq:thermochemistry-H2,eq:thermochemistry-N,eq:thermochemistry-n,eq:thermochemistry-T} the equations governing the evolution of ions, molecules ($\mathrm{CO}$ and \Hmol), photon densities and photon fluxes, with definitions for most symbols provided in the table below.
    Note that, for brevity, we omit sink terms corresponding to the creation of $\mathrm{H_2}$ (sinks in \HI) and of $\mathrm{CO}$ (sinks in \CI{} and $\mathrm{O}$), as well as sources coming from their destruction (of $\mathrm{H}$ for \Hmol{} and of C and O for CO).%
  }\label{tab:thermo-defs}
  \begin{align}
    \frac{\mathrm{d} n_i}{\mathrm{d} t} &= \underbrace{n_{i+1}\left[ n_e \alpha_{i+1}(T) + n_e\alpha_{\mathrm{dust},i+1}(T,Z,N_{j,\mathrm{rad}}) + \sum_{k}^\mathrm{species}n_k \xi_{i+1}(T) \right]}_{\text{Creation by recombination from state $i+1$}} \nonumber &\\
    &+ \underbrace{n_{i-1}\left[n_e \beta_{i-1}(T) + \Gamma_{i-1}(n,N_{j,\mathrm{rad}}) + \zeta_{i-1} + \sum_k^\mathrm{species} n_k \chi_{i-1}(T) \right]}_{\text{Creation by ionisation from state $i-1$}} \nonumber &\\
    &- \underbrace{n_i\left[n_e \alpha_i(T) + n_e \alpha_{\mathrm{dust},i}(T,Z,N_{j,\mathrm{rad}})+\sum_k^\mathrm{species} n_k\xi_i(T)\right]}_{\text{Destruction by recombination to state $i-1$}} \nonumber &\\
    &- \underbrace{n_i\left[n_e \beta_i(T) + \Gamma_i(n, N_{j,\mathrm{rad}}) + \zeta_i + \sum_k^\mathrm{species} n_k\chi_i(T)\right]}_{\text{Destruction by ionisation to state $i+1$}}, \label{eq:thermochemistry-n}& \text{Eqs.~1 and 2, \citealt{katzRAMSESRTZNonequilibriumMetal2022}}\\
    \frac{\mathrm{d} n_\mathrm{CO}}{\mathrm{d}t} & = k_0 n_\mathrm{C\textsc{ii}} n_\mathrm{H_2} \beta - \Gamma_\mathrm{CO}n_\mathrm{CO},\quad \text{with} \quad \beta = \frac{k_1 n_\mathrm{O}/n_\mathrm{H}}{k_1 n_\mathrm{O}/n_\mathrm{H} + \Gamma_{\mathrm{CH_x}}/n_\mathrm{H}},& \text{Eqs.~2 and 3, \citealt{katzPRISMNonEquilibriumMultiphase2022}}  \label{eq:thermochemistry-CO}\\
    \frac{\mathrm{d} n_\mathrm{H_2}}{\mathrm{d}t} & = -C_\mathrm{coll}n_\mathrm{H_2} - (k_\mathrm{UV} + k_\mathrm{LW})n_\mathrm{H_2} + (R_\mathrm{d} + R_\mathrm{p}) n n_{\mathrm{H \textsc{i}}}, &\text{Eq.~9, \citealt{katzInterpretingALMAObservations2017}}  \label{eq:thermochemistry-H2}\\
    \frac{\mathrm{d}E}{\mathrm{d}t} & = \underbrace{\sum_k^\mathrm{processes} \sum_i^\mathrm{species} n_i H_k(T, n, N_{j,\mathrm{rad}})}_{\text{Total heating}} - \underbrace{\sum_k^\mathrm{processes}\sum_i^\mathrm{species}\sum_j^\mathrm{species} n_i n_j C_k(T,n,N_{j,\mathrm{rad}})}_{\text{Total cooling}},\label{eq:thermochemistry-T} &\text{Eq.~4, \citealt{katzPRISMNonEquilibriumMultiphase2022}}\\
    \frac{\mathrm{d} N_{j,\mathrm{rad}}}{\mathrm{d} t}& = -\underbrace{cN_{j,\mathrm{rad}} \sum_i^\mathrm{species} n_i \overline{\sigma}_{i,j}}_{\text{Absorption rate}},  \label{eq:thermochemistry-N} & \text{Eq.~A6, \citealt{rosdahlRAMSESRTRadiationHydrodynamics2013}}\\
    \frac{\mathrm{d} \vb*{F}_{j,\mathrm{rad}}}{\mathrm{d} t} &= -c\vb*{F}_{j,\mathrm{rad}} \sum_i^\mathrm{species} n_i \overline{\sigma}_{i,j}, \label{eq:thermochemistry-F} & \text{Eq.~A8, \citealt{rosdahlRAMSESRTRadiationHydrodynamics2013}}\\
    \text{with} \quad n_e & =\sum_i^\text{species} n_i \times N_\text{positive charges on species $i$}.\label{eq:thermochemistry-e} &
  \end{align}%
  \vspace{-1em}%
  \begin{tabular}{cl}
    \toprule
    \textbf{Symbol} & \textbf{Definition} \\
    \midrule
    $n$   & Total number density\\
    $n_i$ & Number density of ion $i$ (or of molecule $i$) \\
    $n_e$ & Number density of free electrons (derived from $n_i$ to ensure neutrality)\\
    $\alpha_i$ & Total recombination rate coefficient (radiative + dielectronic)\\
    $\alpha_{\mathrm{dust},i}$ & Recombination rate coefficient (on dust)\\
    $T$ & Temperature\\
    $Z$ & Mass fraction of metals (i.e., all elements but hydrogen and helium)\\
    $\xi_i$ & Charge exchange recombination coefficient\\
    $\beta_i$ & Collisional ionisation rate coefficient\\
    $\Gamma_i$ & Photoionisation rate\\
    $\zeta_i$ & Cosmic ray ionisation rate\\
    $\chi_i$ & Charge exchange ionisation coefficient\\
    $\Gamma_\mathrm{CO}$ & Destruction rate of CO (incl.\ UV and cosmic-ray ionizations)\\
    $C_\mathrm{coll}$ & $\mathrm{H_2}$ collisional dissociation rates (with partners $e$, H, He and $\mathrm{H_2}$) \\
    $k_\mathrm{UV,LW}$ & UV and Lyman-Werner photoionisation and photodissociation rates \\
    $R_\mathrm{d,p}$ & $\mathrm{H_2}$ formation rate \emph{via} dust and \emph{via} $\mathrm{H^-}$\\
    $E$ & Energy density\\
    $C_k$ & Cooling rates (process-by-process: primordial, metal-line, dust and molecular cooling)\\
    $H_k$ & Heating rates (process-by-process: photo, photoelectric, cosmic-ray, and $\mathrm{H}_2$ heating)\\
    $N_{j,\mathrm{rad}}$ & Photon number density in frequency bin $j$\\
    $\vb*{F}_{j,\mathrm{rad}}$ & Photon flux vector in frequency bin $j$\\
    $c$ & Speed of light (or reduced speed of light)\\
    $\overline{\sigma}_{i,j}$ & Luminosity-weighted average cross-section for species $i$ in bin $j$\\
    \bottomrule
  \end{tabular}%
\end{table*}

\section{Comparison between non-equilibrium and \PIEUVthinSSnolocal}%
\label{sec:neq_PIE_ionization_level}

\begin{figure*}
    \includegraphics[width=\linewidth]{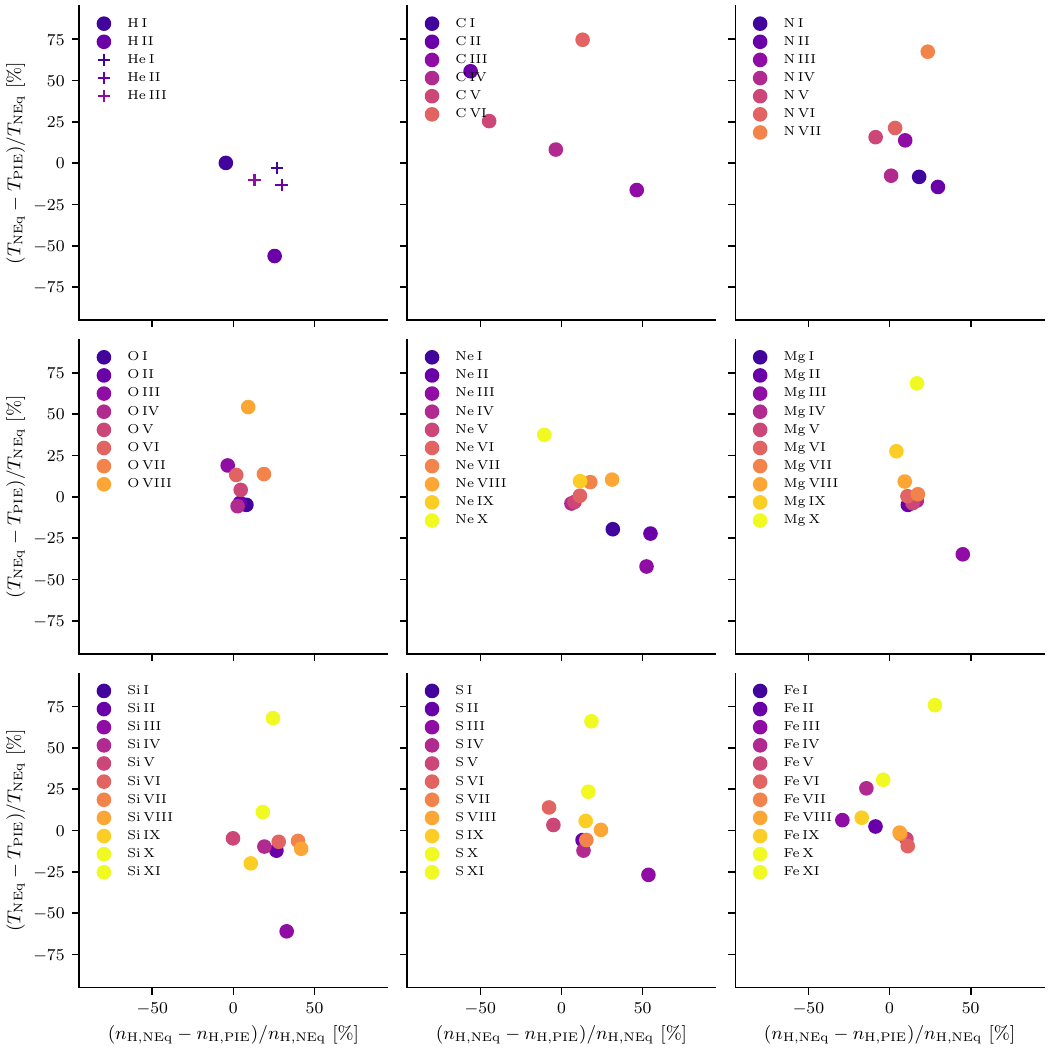}
    \caption{
        Relative difference in the mass-weighted median temperature and density traced by different ions in the CGM of the most massive galaxy at $z=3$ between the non-equilibrium simulation and when assuming \PIEUVthinSSnolocal (abbreviated PIE for brevity).
        Each point is coloured by the ionisation level.
    }\label{fig:avg_T_vs_avg_nH_diff_scatter}
\end{figure*}

We provide additional figures showing the relative difference in the mass-weighted median temperatures and density traced by different ions in the CGM in \cref{fig:avg_T_vs_avg_nH_diff_scatter}; this allows us to show more clearly differences of the order of \SIrange{10}{20}{\percent} that are otherwise non-legible in \cref{fig:avg_T_vs_avg_nH_diff}.

\begin{figure*}
    \includegraphics[width=\linewidth,trim={0 0 {.1\linewidth} 0},clip]{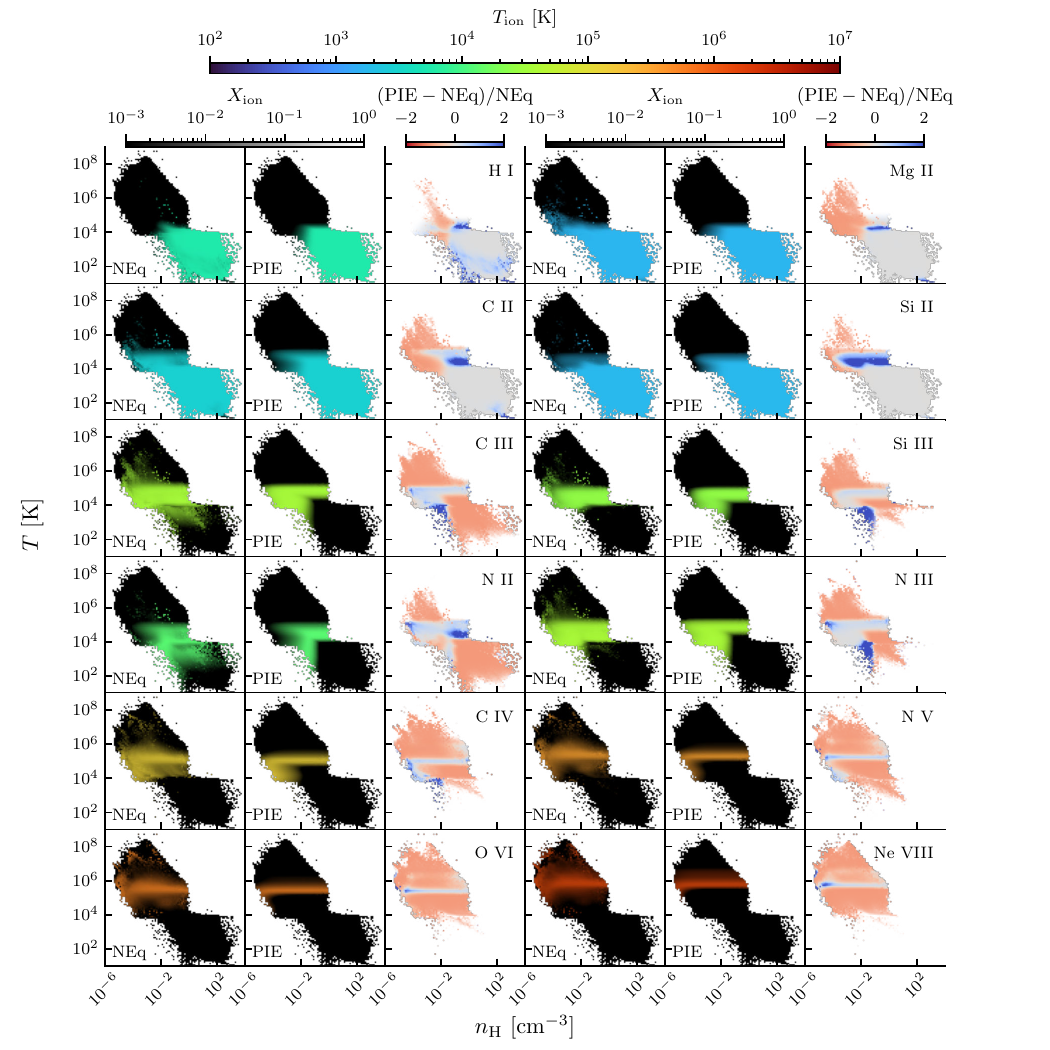}
    \caption{
        Density-temperature diagrams of different ions in the non-equilibrium simulation (first and fourth columns), assuming \PIEUVthinSSnolocal (second and fifth columns, abbreviated PIE for brevity), and their relative difference (third and sixth columns).
        Each panel is shaded by the ionisation fraction and coloured proportionally to the mass-weighted temperature (shades of blue for tracers of the cold CGM, shades of red for tracers of the hot CGM).
    }\label{fig:PIE_vs_NEQ_SS_X_weighted}
\end{figure*}

We also provide the density-temperature diagrams of different ions in the non-equilibrium simulation vs. assuming \PIEUVthinSSnolocal, weighed by ionisation fraction in \cref{fig:PIE_vs_NEQ_SS_X_weighted} (as opposed to weighed by mass fraction in \cref{fig:PIE_vs_NEQ_SS_M_weighted}).

\section{Column densities for additional ions}
\begin{figure*}
    \includegraphics[width=\linewidth]{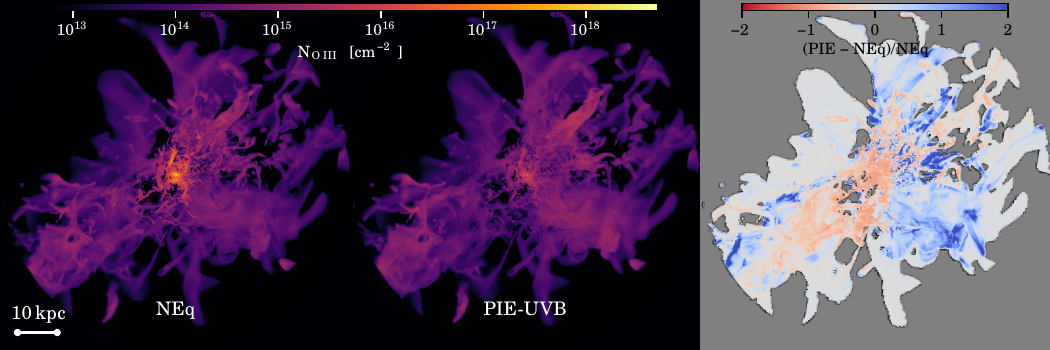}
    \includegraphics[width=\linewidth]{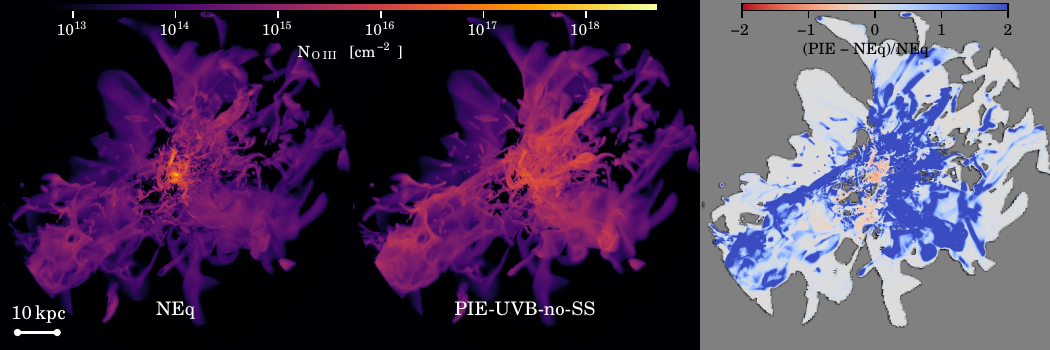}
    \caption{
        \OIII column density maps around the most massive galaxy at $z=4$.
        (Left column) Non-equilibrium simulation (middle column), photoionisation equilibrium,  and (right column), the relative difference.
        We show \PIEUVthinSSnolocal, with self-shielding for all species rather than just those with ionisation energies of $<\SI{13.6}{eV}$ (top row) and \PIEUVthinnolocal, i.e.\ with no self-shielding (bottom row).
    }\label{fig:off_axis_O_III_PIE_variations}
\end{figure*}

Following \cref{fig:off_axis_O_III_SS}, we show in \cref{fig:off_axis_O_III_PIE_variations} column densities of \OIII under the assumption of \PIEUVthinSSnolocal with self-shielding for all species (top row) and under the assumption of \PIEUVthinnolocal (bottom row). Removing self-shielding results in \OII being significantly depleted, resulting in a large increase of \OIII (bottom row of \cref{fig:off_axis_O_III_PIE_variations}).
Conversely, whether \OI is self-shielded (\cref{fig:off_axis_O_III_SS}) or not (\cref{fig:off_axis_O_III_PIE_variations}, top row) does not significantly affect \OIII and higher ionisation states, but does affect \OII.

\begin{figure*}
    \includegraphics[width=\linewidth]{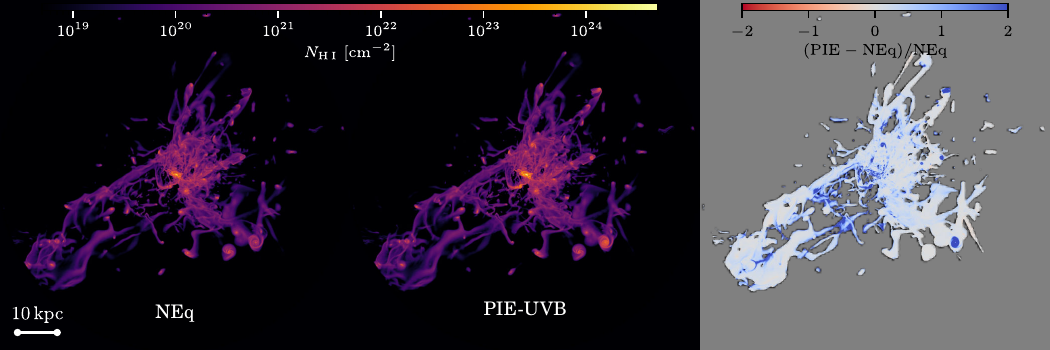}
    \includegraphics[width=\linewidth]{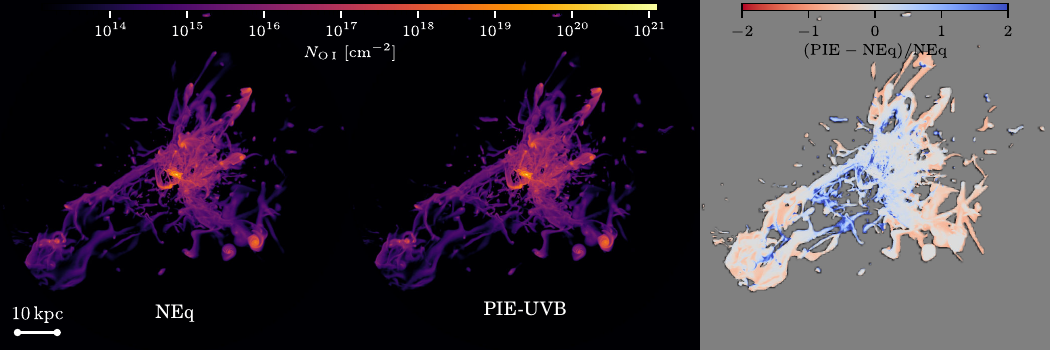}
    \includegraphics[width=\linewidth]{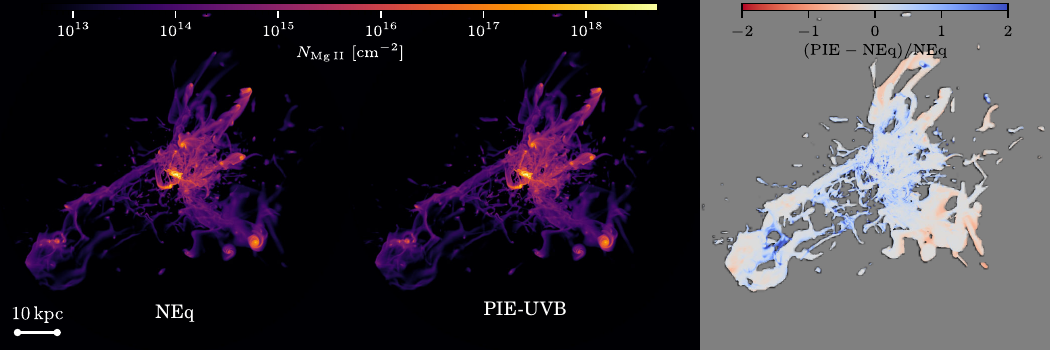}
    \caption{
        \HI, \OI, and \MgII column density maps of the most massive galaxy at $z=4$.
        (Left) Non-equilibrium simulation, (centre) \PIEUVthinSSnolocal and (right), the relative difference between the two.
    }\label{fig:off_axis_H_I_O_I_Mg_II}
\end{figure*}

\begin{figure*}
    \includegraphics[width=\linewidth]{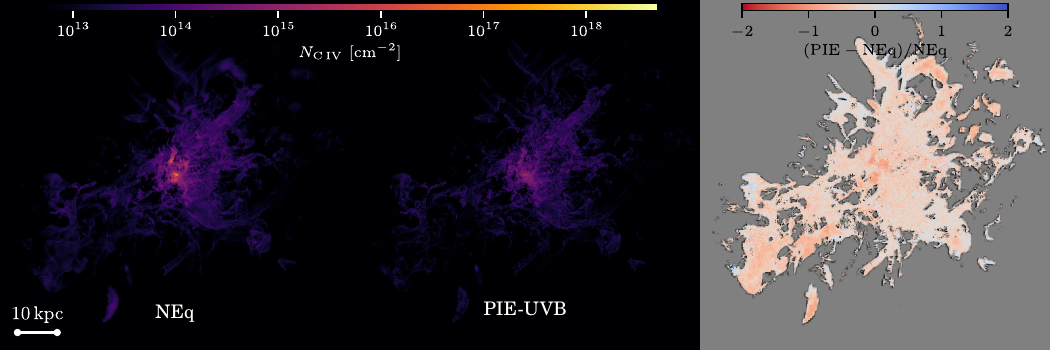}
    \includegraphics[width=\linewidth]{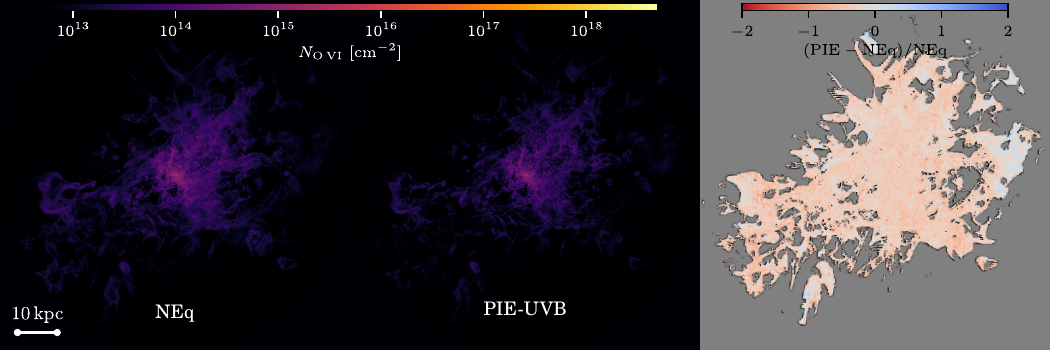}
    \caption{
        (Continued) \CIV, and \OVI column density maps of the most massive galaxy at $z=4$.
        (Left) Non-equilibrium simulation, (centre) \PIEUVthinSSnolocal and (right), the relative difference between the two.
    }\label{fig:off_axis_C_IV_O_VI}
\end{figure*}

Column densities for additional ions (\HI, \OI, \MgII, \CIV, and \OVI) assuming \PIEUVthinSSnolocal vs.\ our non-equilibrium simulation are shown in \cref{fig:off_axis_H_I_O_I_Mg_II,fig:off_axis_C_IV_O_VI}.
These figures illustrate where differences between the non-equilibrium and \PIEUVthinSSnolocal are located spatially, to be compared to their location in $T-n_\mathrm{H}$ space, as shown in \cref{fig:PIE_vs_NEQ_SS_M_weighted}. Notably, observing \HI and \OI, we see that the largest over-abundance in \PIEUVthinSSnolocal compared to the non-equilibrium simulation is found close to Sub-DLAs ($N_{\HI}\sim\SI{e19}{cm^{-2}}$), which are not optically thick: those regions share the same external UVB, but lack the local radiation field in the \PIEUVthinSSnolocal case. However, if the sole origin of this over-abundance was local ionising radiation, we would expect no difference in the abundance of \HI or \OI in DLAs that should be self-shielded from both local and external ionising radiation.

Higher ionisation species, such as \CIV and \OVI, are more abundant in the non-equilibrium simulation than when assuming \PIEUVthinSSnolocal.
For those species, the recombination lag highlighted in \cref{fig:trecomb_tcool_grid} plays a key role in allowing those warm-to-hot gas tracers to survive at a lower temperature than when assuming PIE.


\section{Map of the resolution}
\begin{figure*}
    \includegraphics[width=\linewidth]{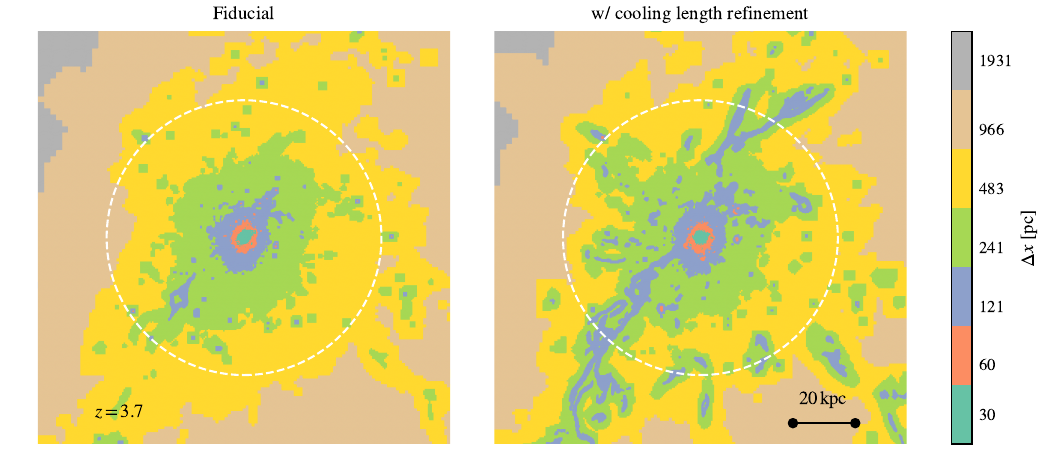}
    \caption{
        Slices of the simulation showing spatial resolution in the fiducial (left), and with additional cooling length refinement (right) around the most massive galaxy at $z=3.7$. The dashed white circle indicates the virial radius of the main halo.
        Most of the additional refinement happens along clumps and filaments in the CGM, but the enhanced resolution clearly extends beyond the virial radius and into the IGM.
    }\label{fig:resolution_plot}
\end{figure*}

We show in \cref{fig:resolution_plot} maps slices of the spatial resolution in the fiducial (left) and with additional cooling length refinement, around the most massive galaxy at $z=3.7$ ($\SI{150}{Myr}$ after $z=4$). The resolution naturally increases in regions that are cooling-unstable, i.e.\ around cosmological cold flows and around clumps. Moreover, we also see additional resolution in the IGM. Note also that similar increases in resolution are obtained around other less-massive haloes.

\end{document}